\documentclass[preprint,amsmath,amssymb,nofootinbib]{revtex4}
\usepackage[dvips]{graphics}
\usepackage{verbatim}
\usepackage{makecell}
\usepackage{booktabs}  
\usepackage[utf8]{inputenc}
\usepackage{multirow}
\usepackage{graphics}
\makeatletter
\renewcommand{\fnum@figure}{Fig.~ \thefigure}
\makeatother
\usepackage{amsfonts}
\usepackage{pstricks}
\usepackage{bm}
\usepackage{amsmath}
\usepackage{amssymb}
\usepackage{color}
\usepackage[all]{xy}
\usepackage{comment}
\usepackage{epstopdf,epsfig}
\def\be{\begin{equation}}
\def\ee{\end{equation}}
\def\bea{\begin{eqnarray}}
\def\eea{\end{eqnarray}}
\usepackage{float}
\usepackage{xcolor}
\usepackage{graphicx}
\usepackage{subcaption}
\usepackage{setspace}
\usepackage{microtype}
\usepackage[
    colorlinks=true,
    linkcolor=blue,     
    citecolor=red,      
]{hyperref}

\begin{document}

\title{Quasi-Homogeneous Thermodynamics and Microscopic Structure of the Quantum-Corrected FLRW Universe}

\author{Carlos E. Romero-Figueroa$^1$ and Hernando Quevedo$^{1,2,3}$}
\email{\raggedright
carlosed.romero@correo.nucleares.unam.mx;
quevedo@nucleares.unam.mx
}
\affiliation{\hspace{1cm}\\ \mbox{$^1$Instituto~de~Ciencias~Nucleares,~Universidad~Nacional~Aut\'onoma~de~M\'exico,}\\ \mbox{AP~70543,~Mexico~City,~Mexico}}
\affiliation{\mbox{$^2$Dipartimento~di~Fisica~and~Icra,~Universit\`a~di~Roma~“La~Sapienza”,~Roma,~Italy}}
\affiliation{\mbox{$^3$Al-Farabi~Kazakh~National~University,~Al-Farabi~av.~71,~050040~Almaty,~Kazakhstan}\\}

\date{\today}

\begin{abstract}
The analysis of phase transitions in cosmological spacetimes shows that their existence requires a time-dependent apparent horizon radius, which in turn implies an equation of state different from that of a dark energy fluid. This condition is not compatible with the simultaneous fulfillment of Hayward's unified gravitational first law and the fundamental thermodynamic equation of the apparent horizon. To solve this problem, we introduce an alternative formulation in which the cosmological horizon is modeled as a quasi-homogeneous thermodynamic system. We apply this approach to the Friedmann-Lema\^itre-Robertson-Walker (FLRW) universe under quantum gravity corrections encoded by the Generalized Uncertainty Principle (GUP),  promote the deformation parameter to a thermodynamic variable, and obtain a consistent thermodynamic description without relying on the usual pressure-volume interpretation. Using Geometrothermodynamics (GTD), we show that fluctuations of the GUP parameter can induce phase transitions closely resembling those of black hole configurations. Finally, we perform a numerical analysis of the behavior of the GTD scalar curvature near the phase transition point, where we find a scaling behavior characterized by the critical exponent close to 1,  independently of the dimension of the equilibrium space. This reveals that quantum gravity corrections not only modify the thermodynamic consistency of cosmological models but also strengthen the notion of thermodynamic universality across gravitational systems. Our findings confirm GTD as a powerful geometric tool to unveil the emergent thermodynamic microstructure of spacetime.\\

\end{abstract}

\maketitle
\tableofcontents
\section{Introduction}

It is remarkable that gravity, usually described through spacetime curvature in Einstein’s field equations, can also be viewed as a thermodynamic system. This deep connection suggests that gravitational dynamics may arise from underlying thermodynamic principles \cite{padmanabhan2010thermodynamical}. Central to this idea is the gravity–thermodynamic conjecture \cite{jacobson1995thermodynamics,padmanabhan2005gravity}, which derives Einstein’s equations from thermodynamic arguments. A key realization is that black holes obey the laws of thermodynamics \cite{bardeen1973four}, initially seen as a geometric analogy \cite{bekenstein1973black}, while Hawking’s discovery of black hole radiation \cite{hawking1974black,hawking1975particle} established their physical temperature, laying the foundation of black hole thermodynamics.
Further developments revealed deep analogies between black hole and ordinary thermodynamics, especially in Anti-de Sitter (AdS) spacetimes, where the cosmological constant acts as a thermodynamic pressure $P=-\Lambda/8\pi G$ and a conjugate volume $V$ naturally appears \cite{kastor2009enthalpy}. In this extended framework, black holes exhibit phase transitions and critical behavior analogous to those of ordinary thermodynamic systems (see \cite{kubizvnak2017black, mann2025black} for detailed discussions). In particular, charged AdS black holes reproduce the characteristic behavior of van der Waals (vdW) fluids \cite{Ladino:2024ned}. This thermodynamic universality manifests itself across many classes of solutions, such as rotating black holes \cite{romero2024extended1,LADINO2025117031}, among others \cite{kubizvnak2017black}.

Black holes are just one example of spacetimes with horizons, for which general relativity (GR) assigns a thermodynamic entropy proportional to the horizon area. The largest known gravitational system, the Universe, should likewise obey thermodynamic laws. Numerous studies of the FLRW universe highlight the central role of the apparent horizon in establishing a consistent thermodynamic framework, as captured by Hayward’s gravitational unified first law \cite{cai2005first,cai2007unified,hayward1994general,hayward1998unified}. Drawing parallels with the thermodynamic phenomena of black holes, the FLRW Universe displays several notable features, including the Hawking temperature \cite{cai2009hawking}, a geometric entropy, and the quasi-local Misner–Sharp (MS) energy \cite{cai2009generalized}.
However, in pure Einstein gravity the cosmological horizon, when regarded as a thermodynamic system, displays a trivial\footnote{Interestingly, recent results \cite{cruz2024new} indicate that holographic-type dark energy can induce nontrivial $P$- $V$ phase transitions in pure Einstein gravity.} phase structure without any critical behavior \cite{abdusattar2022first}. This situation changes markedly in certain modified gravity theories, particularly within the Horndeski class \cite{kong2022p, abdusattar2023phase}, where a clear vdW-like phase structure emerges in the $P$- $V$ plane.
 Building upon this foundation, subsequent studies have examined the phase transitions, critical behavior, and thermodynamic microstructure of the FLRW universe within various modified gravity frameworks \cite{saavedra2025thermodynamic, abdusattar2023insight, abdusattar2023phase}, as well as employing non-additive entropies, such as the Kaniadakis entropy \cite{housset2024cosmological, sheykhi2024corrections} and the Barrow entropy \cite{sheykhi2021barrow}. These studies indicate that thermodynamic phase transitions may also emerge in FLRW universe models formulated beyond the framework of classical GR. In this context, a wide range of proposals have been developed to extend or modify GR, among which quantum gravity (QG) stands out as a particularly prominent avenue.
 
 In particular, several approaches within QG, together with developments in black hole physics \cite{maggiore1993algebraic, gross1987high} and string theory \cite{amati1989can, konishi1990minimum}, consistently predict the existence of a fundamental minimal length scale \cite{capozziello2000generalized, maggiore1993algebraic}. Theoretical models that incorporate such a minimal length scale and/or a maximal momentum are commonly referred to as frameworks based on the generalized uncertainty principle (GUP) (see \cite{tawfik2015review} and references therein). The postulation of a fundamental minimal length is inherently incompatible with ordinary quantum mechanics, which assumes a continuous spacetime structure down to arbitrarily small scales. Consequently, it becomes necessary to generalize the Heisenberg uncertainty principle to accommodate this minimal length. These generalizations frequently arise from modifications of the canonical commutation relations \cite{ali2009discreteness, kempf1995hilbert}, and in some formulations of string theory, they are also associated with the emergence of non-commutative geometries \cite{seiberg1999string, konechny2002introduction}. Interestingly, as shown in \cite{nozari2007thermodynamics}, GUP and space non-commutativity can be regarded as conceptually similar, since both lead to identical thermodynamic corrections in black hole thermodynamics. Furthermore, a GUP can be formulated without altering the underlying commutation relations. In this work, we focus on investigating the impact of the GUP parameter on the thermodynamics of the FLRW universe. This model is based on a modified Schwarzschild black hole entropy that incorporates back-reaction effects from quantum tunneling \cite{banerjee2008quantum}. Within this framework, a heuristic derivation of the GUP is proposed in \cite{du2022new} and can be expressed as
\begin{equation}
\Delta x \Delta p \geq \frac{\hbar}{2} \left(\frac{1}{1 + 16\beta/(\Delta x)^2}\right).\label{GUP}
\end{equation}
In this relation, $\Delta x$ and $\Delta p$ represent the uncertainties in position and momentum, respectively. Moreover, $\beta = \beta_0 \ell_p^2$ is the deformation or GUP parameter, with $\beta_0$ arising from the modified surface gravity \cite{york1985black}, and $\ell_p$ denotes the Planck length. For this GUP model, recent works have established bounds on the parameter $\beta_0$. For instance, in \cite{luo2023gravitational} using the baryon asymmetry factor and observational data, it was found that $
|\beta_0| \lesssim 10^9$, while constraints from Big Bang nucleosynthesis give \cite{luo2025new} $
|\beta_0| \lesssim 10^{84}$. This formulation of the uncertainty principle consistently yields a fixed and unified minimal length scale, given by $\Delta x_{\text{min}} = 4 \sqrt{|\beta|}$, independent of the sign of the deformation parameter. This feature is not shared by other GUP proposals, such as those in \cite{kempf1995hilbert}, \cite{ali2009discreteness}, \cite{pedram2012higher}, and \cite{chung2019new}, where the minimal length may depend on the sign or structure of the deformation function. It is particularly relevant, as shown in~\cite{feng2024phase}, that the parameter $\beta$ modifies the Friedmann equations and, consequently, the corresponding equation of state, given by
\begin{equation}
P = T_A\left[\left(\frac{\pi}{6V}\right)^{1/3} + \frac{8\pi\beta}{3V}\right]
+ \frac{1}{V^{4/3}} \, \frac{3\times6^{1/3}V^{2/3} - 4\times(6\pi)^{2/3}\beta}{36\pi^{1/3}}, \label{Feng_EOS}
\end{equation}
where $P \equiv W$ denotes the work density, as defined below in 
Eq.~\eqref{work densi}, $V = 4\pi R_A^3/3$ represents the volume enclosed by the cosmological horizon, and $T_A$ is the geometric temperature of the horizon given below in Eq. \eqref{surf_horizon}. In~\cite{feng2024phase}, both $P$ and $V$ are treated as fluctuating thermodynamic variables, while $\beta$ is considered a fixed parameter. The authors showed that under these assumptions, $\beta$ can induce phase transitions in the modified cosmological model. Additionally, employing Ruppeiner’s geometric thermodynamics~\cite{ruppeiner1979thermodynamics}, the nature of the phase transition and the associated microscopic properties were also investigated. However, Eq.~\eqref{Feng_EOS} is valid only when $\dot{R}_A \neq 0$. As we demonstrate in Sec.~\ref{funda_sect}, this assumption is incompatible with the simultaneous fulfillment of the unified gravitational first law and the thermodynamic first law derived from the system's fundamental equation. To resolve this inconsistency,  we propose an alternative formulation that preserves consistency with Euler scaling relations~\cite{callen1998thermodynamics}. In this approach, the deformation parameter $\beta$ is interpreted as an effective fluctuating thermodynamic variable. This perspective naturally leads to viewing the cosmological horizon as a quasi-homogeneous thermodynamic system (see~\cite{quevedo2019quasi} for a detailed review of quasi-homogeneous functions). In this manner, we develop a consistent model in which the GUP parameter can induce phase transitions while remaining compatible with thermodynamic scaling arguments.\\\\ This paper is organized as follows: In Sec.~\ref{FLRW section}, we formulate the notion of quasi-homogeneous thermodynamics for the GUP-corrected FLRW universe and derive the corresponding fundamental equation consistent with the unified first law. Section~\ref{GTD} introduces the GTD formalism and applies it to both Einstein gravity and the GUP-modified FLRW universe. In Sec.~\ref{micro}, we investigate the thermodynamic microstructure of the system through the analysis of the GTD scalar curvature and its critical behavior. Finally, Sec.~\ref{conc} summarizes our main conclusions and outlines possible directions for future research.

\section{The GUP-FLRW horizon and thermodynamics}
\label{FLRW section}
First, we review the fundamental properties of the FLRW solution in Einstein gravity corrected by a GUP parameter. Consider a 4-dimensional spherically symmetric spacetime described by the line element
\begin{equation}
ds^{2} = h_{ab}\,dx^{a}dx^{b} + R^{2}\left(d\theta^{2} + \sin^{2}\theta\, d\phi^{2}\right), \label{metricFLRW}
\end{equation}

\noindent with
\begin{equation}
h_{ab} = \mathrm{diag}\!\left(-1,\;\frac{a^{2}(t)}{1 - k r^{2}}\right), 
\qquad
x^{0} = t, \; x^{1} = r, 
\qquad
R(t,r) = a(t)r,
\end{equation}

\noindent where $k=0, \pm 1$ denotes the curvature of the spatial sections. The FLRW  metric (\ref{metricFLRW}) possesses a marginally trapped
null surface, characterized by a vanishing expansion. This surface coincides with the cosmological apparent horizon and satisfies the condition \cite{cai2005first,cai2007unified}
\begin{equation}
h^{ab} \, \frac{\partial R}{\partial x^{a}} \, \frac{\partial R}{\partial x^{b}} = 0
\end{equation}
whose solution reads

\begin{equation}
R_A = \frac{1}{\sqrt{H^2 + \tfrac{k}{a^2}}}, \label{Rh horizon}
\end{equation}
where $H = \dot{a}/a$ denotes the Hubble parameter, which determines the radius 
of the apparent horizon. By definition, the apparent horizon, being a trapped 
surface, has an associated surface gravity. The surface gravity $\kappa$ 
and the corresponding horizon temperature $T_A$ are then given by \cite{cai2005first}
\begin{equation}
   \kappa = \frac{1}{2\sqrt{-h}} \, \frac{\partial}{\partial x^a} 
   \left( \sqrt{-h} \, h^{ab} \frac{\partial R}{\partial x^b} \right), 
   \qquad  
T_A = \frac{\kappa_A}{2\pi}
    = \frac{1}{2\pi R_A} \left( 1 - \frac{1}{2} \frac{\dot{R}_A}{H R_A} \right),
   \label{surf_horizon}
\end{equation}
where $h$ denotes the determinant of the metric $h_{ab}$. Notice that for very slowly changing apparent horizon, the temperature reduces to $T_A = 1\big/2 \pi R_A$,
an expression that resembles the temperature of a spherically symmetric black hole with horizon radius $R_A$. In addition, for the metric \eqref{metricFLRW}, 
the MS energy is defined as \cite{misner1964relativistic,cai2009generalized}
\begin{equation}
    E = \frac{R}{2} \left( 1 - h^{ab} \frac{\partial R}{\partial x^a} \frac{\partial R}{\partial x^b} \right). \label{MS energy}
\end{equation}
Similarly to $\kappa$, the MS energy is a purely geometric quantity, determined entirely by the spacetime metric. When evaluated using the field equations of Einstein gravity sourced by a perfect fluid, $R^{\mu\nu} - \frac{1}{2} g^{\mu\nu} R = 8 \pi T^{\mu\nu}$, Eq.~\eqref{MS energy} reduces to $E = \rho V$, representing the energy contained within a sphere of radius $R$. In modified gravity theories, Eq.~\eqref{MS energy} is generalized to incorporate higher-order curvature effects, quantum corrections, or extensions such as $f(R)$ gravity. In such cases, it can be written as $E_{eff} = \rho V$, where $E_{ eff}$ denotes the effective energy, incorporating both the matter and the gravitational contributions \cite{cai2009generalized}.
 Furthermore, by introducing the work density $W$ and the energy-supply vector $\Psi$, which are invariants constructed from the energy stress tensor 
\begin{equation}
  W \equiv -\frac{1}{2} T_{ab} h^{ab}, \quad 
 \Psi \equiv \Psi_a dx^a=
\left(T^{\ b}_{a} \frac{\partial R}{\partial x^b} + W \frac{\partial R}{\partial x^a} \, \right)dx^a; \label{work densi}
\end{equation}
for a spherically symmetric spacetime, the Einstein equations can be written in the following form
\begin{equation}
    d\Tilde{E}_{eff} = \Tilde{A} \Tilde{\Psi} + \Tilde{W} d\Tilde{V}, \label{unified law}
\end{equation}
where in 4 dimensions, $\Tilde{A} = 4\pi R^2$ is the area of a sphere with radius $R$, and 
$\Tilde{V} = 4\pi R^3/3$ is its corresponding volume. 
Eq.~\eqref{unified law}, first proposed as a general definition of black hole dynamics on trapping horizons in Einstein theory, is known as 
Hayward’s unified first law for gravitational systems~\cite{hayward1998unified, hayward1994general}. 
By projecting Eq.~\eqref{unified law} onto the apparent cosmological horizon, Eq.~\eqref{Rh horizon}, and assuming that the horizon behaves as a thermodynamic system whose geometric entropy is a function of the area, $S = f(A)/4$, one obtains the first law of thermodynamics

\begin{equation}
    dE_{eff}=-T_AdS+WdV \label{first law2}.
\end{equation}
From the above equation, we identify the thermodynamic pressure as $P \equiv W$, and the Clausius relation in Eq.~\eqref{first law2} 
indicates that the thermodynamic internal energy $U$ corresponds to the 
negative of $E_{eff}=\rho V$ \cite{kong2022p, sanchez2023thermodynamics}. To derive the GUP-corrected Friedmann's equations, we start with the modified geometric entropy\footnote{We use units in which $\ell_p = 1$ and the Boltzmann constant $k_B = 1$.
}
 proposed in \cite{feng2024phase,luo2023gravitational}  
\begin{equation}
S = \frac{A}{4} + 4\pi\beta \ln \left( \frac{A}{\sigma} \right), \label{eqSGup}
\end{equation}
where \(A = 4\pi R_A^2\) denotes the area of the apparent horizon, \(\beta\) is the GUP parameter introduced in Eq.~\eqref{GUP}, and \(\sigma\) is a positive integration constant with dimensions of area. Note that logarithmic corrections to the entropy also appear in loop quantum gravity approaches~\cite{rovelli1996black} and in studies of corrections induced by thermal or quantum fluctuations~\cite{cai2008corrected}. Furthermore, in the limit \(\beta \to 0\), one recovers the standard Bekenstein--Hawking entropy, \(S = A/4\). Next, assuming that the spacetime described by Eq.~\eqref{metricFLRW} is sourced by a perfect fluid with energy--momentum tensor

\begin{equation}
T^{\mu\nu} = (\rho+ p)\,u^{\mu}u^{\nu} + p\,g^{\mu\nu},
\end{equation}
where $\rho$ and $p$ denote the fluid energy density and pressure, respectively, and $u^\nu$ is its four-velocity. Then, according to Eq.~\eqref{work densi}, the work density for a perfect fluid is given by

\begin{equation}
    W = \frac{\rho-p}{2}.\label{Work_pfluid}
\end{equation}

Using Eqs.~\eqref{eqSGup}, \eqref{surf_horizon}, and \eqref{Work_pfluid}, the right-hand side of the first law can be expressed as

\begin{align}
T_A \, dS &= - \frac{1}{8 \pi R_A} \left(1 - \frac{\dot{R}_A}{2 H R_A} \right) \left( 1 + \frac{16 \pi \beta}{A} \right) dA, \label{TDS} \\
W \, dV &= 2 \pi R_A^2 \left(\rho - p\right) \, dV,
\end{align}  
while the left-hand side reads  
\begin{equation}
   dE_{\text{eff}} = d(\rho V) = A \rho \, dR_A + V \, d\rho.
\end{equation}  
From Eq.~\eqref{Rh horizon}, the apparent horizon satisfies  
\begin{equation}
   \dot{R}_A = - H R_A^3 \left( \dot{H} - \frac{k}{a^2} \right). \label{derivaR}
\end{equation}
Substituting Eq.~\eqref{derivaR} into Eq.~\eqref{TDS}, one obtains the following relations

\begin{align}
-4\pi (\rho + p) &= \left( 1 + \frac{16 \pi \beta}{A} \right) \left( \dot{H} - \frac{k}{a^2}\right), \\
\frac{8\pi}{3} \rho &= -4 \pi \int \left( 1 + \frac{16 \pi \beta}{A} \right) \frac{dA}{A^2}  \label{FLRW2}+ \text{constant}.
\end{align}  
The integration constant in Eq.~\eqref{FLRW2} can be calibrated as a cosmological constant \cite{das2022baryon}. In the present work, we ignore it for simplicity in order to focus on the effects of the GUP. Finally, inserting Eq. \eqref{Rh horizon} into the above equations, we obtain the GUP-corrected Friedmann's equations

\begin{align}
\left[1 + 4\beta \left(H^2 + \frac{k}{a^2}\right)\right] \left(\dot{H} - \frac{k}{a^2}\right) &= - 4 \pi (\rho + p), \label{Fried_eq2} \\
H^2 + \frac{k}{a^2} + 2\beta \left(H^2 + \frac{k}{a^2}\right)^2 &= \frac{8 \pi}{3} \rho. \label{fried_equ1}
\end{align}
Interestingly, these equations share the same structural form as those obtained in scalar–tensor theories within Horndeski gravity \cite{kong2022p}, as well as in four-dimensional regularized Einstein–Gauss–Bonnet (EGB) gravity \cite{saavedra2025thermodynamic,feng2021theoretical,sanchez2025thermodynamics}. Next, we analyze the main thermodynamic properties of the GUP-FLRW universe by treating the cosmological horizon as a thermodynamic system.

\subsection{Fundamental equation of  the GUP-FLRW universe}
\label{funda_sect}
In classical thermodynamics, any system can be fully characterized by a fundamental equation, from which all thermodynamic properties can be derived \cite{callen1998thermodynamics}. To begin, we solve for \(R_A\) using Eq.~\eqref{Rh horizon} and Eq.~\eqref{fried_equ1}, which yields

\begin{equation}
{R_A}^\pm = \frac{\sqrt{3 \pm \sqrt{9 + 192\, \beta \pi \rho}}}{4 \sqrt{\pi \rho}}, \label{R hor2} 
\end{equation}
where only \({R_A}^+\) has physical significance, since in the limit \(\beta \rightarrow 0\) it recovers the correct horizon radius in Einstein gravity, \(R_E = \sqrt{3/8\pi \rho}\) \cite{sanchez2023thermodynamics}. Therefore, substituting \({R_A}^+\) into the entropy expression \eqref{eqSGup} and using \(\rho = U/V\), we obtain the fundamental equation in the entropy representation

\begin{equation}
\widetilde{S} = \frac{ V\left( 3 + \sqrt{9 + 192 \pi U\beta/V} \right)}{16 E} 
+ 4  \pi\beta \, \ln \left[\frac{3 + \sqrt{9 + 192  \pi U\beta/V}}{4\sigma U/V} \right] \, .\label{funda_opcion2}
\end{equation}
In addition, using Eq. \eqref{funda_opcion2}, the thermodynamic first law can be expressed as 
 \begin{equation}
     dU=-\widetilde{T}d \widetilde{S}+PdV, \label{UFL}
 \end{equation}
where the temperature \(\widetilde{T}\) and the work density \(W\) are treated as equations of state, given by
\begin{equation}
    \widetilde{T}=- \left( \frac{\partial \widetilde{S}}{\partial U} \right)^{-1}=\frac{32 \beta \pi U + 3 V - V \sqrt{9 + \frac{192 \beta \pi U}{V}}}{64 \beta^2 \pi^2} \, , \qquad P=\widetilde{T} \left( \frac{\partial \widetilde{S}}{\partial V} \right)=\frac{U}{V}=\rho. \label{Work dens2}
\end{equation}
In the limit $\beta \rightarrow 0$, the fundamental equation \eqref{funda_opcion2} reduces to that of Einstein gravity, $\widetilde{S}_E = 3V/8U$, and the corresponding temperature becomes $\widetilde{T}_E = 8U^2/3V$~\cite{sanchez2023thermodynamics}. Notice that, to be consistent with the unified first law \eqref{first law2}, we must identify $P \equiv W$, which, according to the fundamental equation, gives $W = \rho$ (Eq.~\eqref{Work dens2}), whereas for a perfect fluid \eqref{Work_pfluid} it yields $W = (\rho - p)/2$. Consequently, the simultaneous fulfillment of these two independent conditions requires that the horizon obeys a barotropic equation of state $p = w \rho$ with $w = -1$, characteristic of a dark energy fluid. One of the most significant implications of this result follows directly from Eq.~\eqref{derivaR} upon inserting Eq.~\eqref{Rh horizon} into Eq.~\eqref{Fried_eq2}
\begin{equation}
    \dot{R}_A = 4 \pi \rho H R_A^3 (w+1) \left( 1 + \frac{4 \beta}{R_A^2} \right)^{-1}, \label{dotR}
\end{equation}
which vanishes for \( w = -1 \). Since \(\dot{S} \propto \dot{R}_A \), it follows that the evolution of the horizon is an isentropic process. Remarkably, this result also holds in both the EGB and Einstein theories~\cite{sanchez2023thermodynamics,sanchez2025thermodynamics}, highlighting a thermodynamic universality of the horizon. It would be of considerable interest to examine whether this universality persists in models featuring non-additive entropies \cite{housset2024cosmological}, as well as within the framework of scalar-tensor gravity theories \cite{kong2022p,abdusattar2022first,abdusattar2023phase}. These results show that Eq.~\eqref{Feng_EOS}, which explicitly assumes $\dot{R}_A \neq 0$, conflicts with the first law derived from the fundamental equation Eq.~\eqref{funda_opcion2}. This issue is closely related to the argument presented in \cite{kong2025effective}, where it is pointed out that, since both \(S\) and \(V\) are functions of the apparent horizon radius \(R_A\), heat and work are not independent, in contrast to what occurs in standard thermodynamic systems. To address this peculiarity, the authors propose effective definitions of pressure and temperature in order to restore the standard Clausius relation. For further details, see \cite{kong2025effective}. Furthermore, if we instead use the equation of state $\widetilde{T}(U,V)$ given in Eq.~\eqref{Work dens2}, it can be shown that the system exhibits a trivial phase structure, similar to that encountered in EGB gravity \cite{sanchez2025thermodynamics}.

\begin{figure}[H]
\begin{minipage}[t]{0.4\linewidth}
 \centering
\hspace{1cm}
\includegraphics[height=6cm]{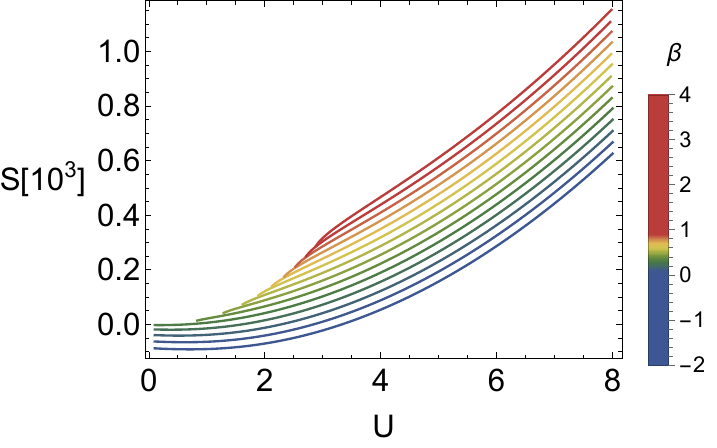}
 (a)\hspace{10cm}
\end{minipage}%
\hfill%
\begin{minipage}[t]{0.4\linewidth}
 \centering
\hspace{1cm}
\includegraphics[height=6cm]
{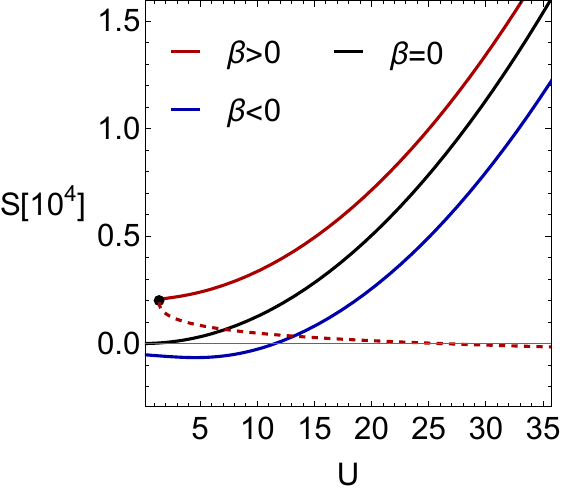}
(b)\hspace{12cm}
\end{minipage}%
\hfill%

\caption{(a) Entropy profile for various values of $\beta$ with $\sigma = 1$. (b) $S$--$U$ diagram for $\sigma = 1$, showing curves corresponding to $\beta = 1$ (red), $\beta = 0$ (black) and $\beta = -10$ (blue). The dashed red curve represents the $R_A^{-}$ branch given by Eq.~\eqref{eqr}, associated with negative temperature $(T < 0)$.}
\label{fig:1}
\end{figure}

\begin{figure}[H]
\begin{minipage}[t]{0.4\linewidth}
 \centering
\hspace{1cm}
\includegraphics[height=6cm]
{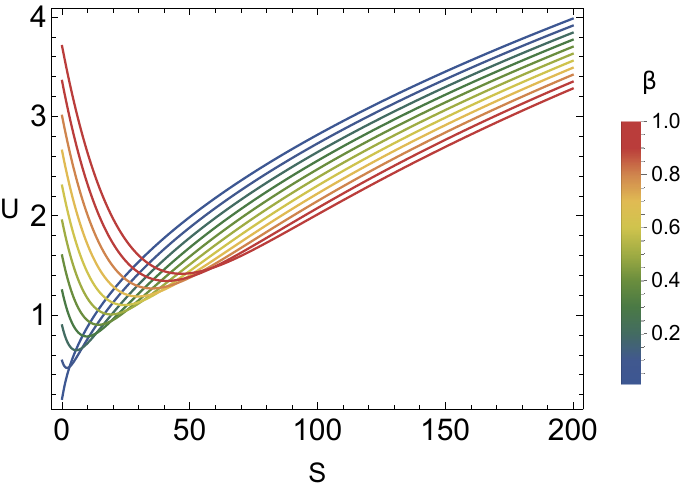}
 (a)\hspace{10cm}
\end{minipage}%
\hfill%
\begin{minipage}[t]{0.35\linewidth}
 \centering
\hspace{1cm}
\includegraphics[height=6cm]{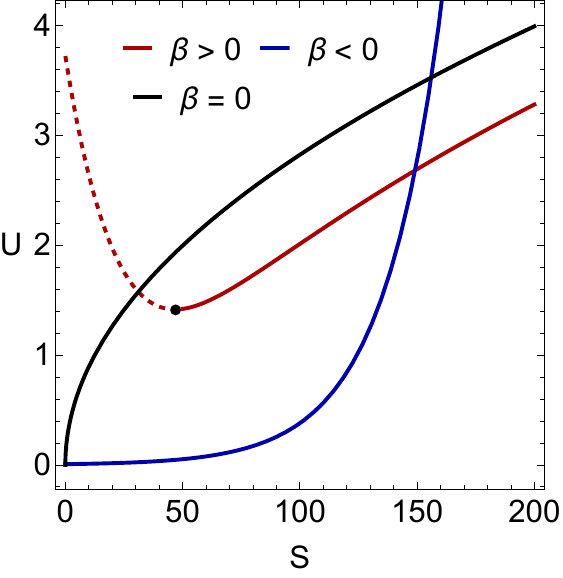}
(b)\hspace{12cm}
\end{minipage}%
\hfill%
\caption{ (a) Energy profile for different $\beta\geq0$ and $\sigma=1$. (b) $U$--$S$ plot for $\sigma=1$. The dashed red curve highlights regions where $T < 0$. A nonzero $U_{\min}$ emerges for $\beta > 0$.}
\label{fig:energy}
\end{figure}

Nevertheless, our aim is to address the question of whether the GUP parameter can trigger phase transitions while remaining consistent with Euler scaling relations. Building on the results of our previous studies on AdS black holes \cite{LADINO2025117031,Ladino:2024ned,romero2024extended1}, in which the thermodynamic properties of black holes can be analyzed without resorting to the concepts of volume and pressure, we propose to study the GUP-FLRW horizon, treating $\beta$ as a thermodynamic parameter and identifying the thermodynamic energy \(U\) with the effective energy evaluated at the horizon, which, according to Eq.~\eqref{fried_equ1} and Eq.~\eqref{Rh horizon}, can be expressed as

\begin{equation}
   U\equiv \rho V\big|_{R=R_A}=\frac{ R_A^2 + 2\beta}{2R_A}. \label{eq2}
\end{equation} 
From Eq. \eqref{eq2} we solve for $R_A(U,\beta)$ 
\begin{equation}
    R_A^{\pm}=U \pm \sqrt{ U^2-2\beta}.\label{eqr}
\end{equation}
The apparent horizon represents a positive length scale \( R_A > 0 \). However, from a theoretical standpoint, the deformation parameter \( \beta \) is not restricted to positive values; negative values are also admissible \cite{luo2023gravitational,luo2025new}. Consequently, from Eq.~\eqref{eqr}, we obtain two distinct solutions. For \(0 < \sqrt{2\beta} \leq U\), both \(R_A^{\pm} > 0\), whereas for \(\beta \leq 0\), we find \(R_A^{+} > 0\) while \(R_A^{-} < 0\). We discard \(R_A^{-}\) since it can be shown that this solution is nonphysical, yielding only negative temperatures (see Fig.\ref{fig:1}(b)). Then, by combining Eq.~\eqref{eqSGup} with Eq.~\eqref{eqr}, the entropy can be written as

\begin{equation}
S(U,\beta,\sigma)= \pi \left ( U + \sqrt{ U^2-2\beta}\right)^2 + 4\pi\beta \ln\left[\frac{4\pi  \left(U + \sqrt{ U^2-2\beta}\right)^2}{\sigma} \right]. \label{funda} 
\end{equation}
Thus, the entropy can be expressed solely in terms of $U$, $\beta$, and $\sigma$ via Eq.~\eqref{funda}. This expression can be inverted to yield the energy representation as

\begin{equation}
U(S,\beta,\sigma) = \frac{\sqrt{\beta} \left( 1 + 2 K \right)}{2 \sqrt{K}}.\label{funda_U}
\end{equation} 
Here, \( K = \textit{ProductLog}\left[e^{S / 4\pi \beta} \left( \sigma / 16\pi \beta \right) \right] \) denotes the product logarithm function, defined as the multivalued inverse of \( f(x) = x e^x \), with \( x \in \mathbb{C} \). The energy representation will be particularly useful in Sec.~\ref{seccion_global}, where we analyze the global thermodynamic stability. For the moment, however, we focus our analysis on the entropy representation. For $\beta\neq 0$, logarithmic corrections can yield negative entropy, with $\sigma$ setting its minimum value. This defines the critical values $\sigma_{\pm} = \pm 8\pi\beta e^{\pm 1/2}$. For $\beta>0$, the minimum internal energy is $U_{\min} = \sqrt{2\beta}$. If $\sigma > \sigma_+$, entropy becomes negative in some region; for $0 < \sigma \le \sigma_+$, $S_{\min}\ge0$. For $\beta < 0$, we have $U_{\min} = 0$, and the entropy becomes negative in the range $0 < \sigma < \sigma_-$, while $S_{\min} \ge 0$ for $\sigma \ge \sigma_-$. This behavior is illustrated in Fig.~\ref{fig:1}. Moreover, Fig.~\ref{fig:energy} shows the internal energy, which for $\beta > 0$ resembles the remnant mass characteristic of GUP-corrected black holes; see, for instance, Ref.~\cite{liu2025thermodynamic}.

Furthermore, an inspection of Eq.~(\ref{funda}) reveals that, to remain consistent with thermodynamic scaling arguments, the deformation parameter $\beta$ and the integration constant $\sigma$ must be treated as independent thermodynamic variables~\cite{quevedo2019quasi}. Thus, the entropy is a quasi-homogeneous function of arbitrary degree $\nu_S$, i.e.,
\begin{equation}
S(\lambda^{\nu_U} U, \lambda^{\nu_\beta} \beta, \lambda^{\nu_\sigma} \sigma) = \lambda^{\nu_S} S(U,\beta,\sigma) \label{scal1}
\end{equation}
if the conditions
\begin{equation}
    \nu_U = \frac{1}{2}\nu_\beta, \quad \nu_\beta = \nu_S, \quad \nu_\sigma = \nu_\beta, \label{escal2}
\end{equation}
are imposed \cite{quevedo2019quasi}. In the entropy representation, we identify the thermodynamic coordinate space  
\( q^a = \{U, \beta, \sigma\} \), 
and define the corresponding dual thermodynamic variables 
\( p_a = \{1/T, B/T, \Sigma/T\} \) 

\begin{equation}
\frac{1}{T} = \frac{\partial S(U,\beta,\sigma)}{\partial U}, \quad  \frac{B}{T} = \frac{\partial S(U,\beta,\sigma)}{\partial \beta}, \quad \frac{\Sigma}{T} = \frac{\partial S(U,\beta,\sigma)}{\partial \sigma}.\label{duals}
\end{equation}
With these conditions, it is then trivial to check that the quasi-homogeneous Euler identity holds
\begin{equation}
    S=\frac{U}{2 T}+\frac{B}{T} \beta+\frac{\Sigma}{T} \sigma.\label{euler}
\end{equation}
We note that the unified first law, Eq.~\eqref{first law2}, in its differential form is always satisfied for infinitesimal variations of the effective energy. Nonetheless, in the context of FLRW cosmology, Hayward's unified first law generally does not admit an integral formulation, as such a relation requires the quasi-homogeneity of the thermodynamic potential, which is typically absent in a dynamical FLRW universe. Therefore, by employing Eq.~\eqref{euler}, one can derive a thermodynamic first law for the GUP-corrected case that is consistent with the scaling relations, as

\begin{equation}
    dS = p_a \, dq^a = \frac{1}{T} \, dU + \frac{B}{T} \, d\beta + \frac{\Sigma}{T} d\sigma, \label{first_law}
\end{equation}
provided that the generalized Gibbs–Duhem relation is satisfied~\cite{bravetti2017zeroth,quevedo2019quasi}
\begin{equation}
    \beta \, d\left(\frac{B}{T}\right) + \sigma \, d\left(\frac{\Sigma}{T}\right )+ \frac{1}{2} U \, d\left(\frac{1}{T}\right) - \frac{1}{2T} \, dU = 0. \label{gi-duh}
\end{equation}
It is important to highlight that the Gibbs-Duhem relation \eqref{gi-duh} deviates from the standard form $q_adp^a=0$ due to the quasi-homogeneous nature of the entropy function \cite{bravetti2017zeroth}. 

We now turn to the analysis of the thermodynamic temperature defined through Eq.~\eqref{duals} as the dual of the energy 

\begin{equation}
    T(U,\beta)\equiv \left( \frac{\partial S}{\partial U}\right)^{-1}=\frac{\sqrt{ U^2-2\beta}}{4\pi \left[ \beta + U \left(U + \sqrt{U^2-2\beta} \right) \right]}.\label{temperature}
\end{equation}
Notably, Eq.~\eqref{temperature} differs from the geometric temperature \( T_A \), defined through the surface gravity in Eq.~\eqref{surf_horizon}. An analogous distinction between two independent temperatures also appears in inflationary models within the GTD framework, as discussed in~\cite{anaya2024thermodynamic}. To compare both temperatures, we express \( T_A \) in terms of the thermodynamic variables using Eqs.~\eqref{dotR} and~\eqref{eq2}, namely,
\begin{equation}
T_A(U,\beta, \omega)=
-\frac{2\beta + U\!\left(U + \sqrt{U^{2} - 2\beta}\right)\!\left(-1 + 3\omega\right)}
{4\pi \left(U + \sqrt{U^{2} - 2\beta}\right)\!\left(-\beta + 2U\left(U + \sqrt{U^{2} - 2\beta}\right)\right)}\,,
\label{geometric tempe}
\end{equation}
which, in general, differs from the thermodynamic temperature \( T \) given in Eq.~\eqref{temperature}. Both temperatures become proportional only in the case \( \beta = 0 \), where $T_A=\frac{1-3\omega}{2}T$,  with 
$T_A = 2 T$ for $\omega=-1$.
 A natural explanation for this discrepancy lies in the fact that \( T_A \), defined via the surface gravity, is independent of the equations of motion, since it is constructed purely from the metric \( h_{ab} \). In contrast, the temperature given by Eq.~\eqref{temperature}, defined as the derivative of the entropy, explicitly depends on the underlying gravitational theory. In particular, the presence of $\beta$ modifies the entropy, making it no longer proportional to one-quarter of the horizon area.
\begin{figure}[H]
\centering
\begin{minipage}[t]{0.46\linewidth}
    \centering
    \includegraphics[width=1.1\textwidth]{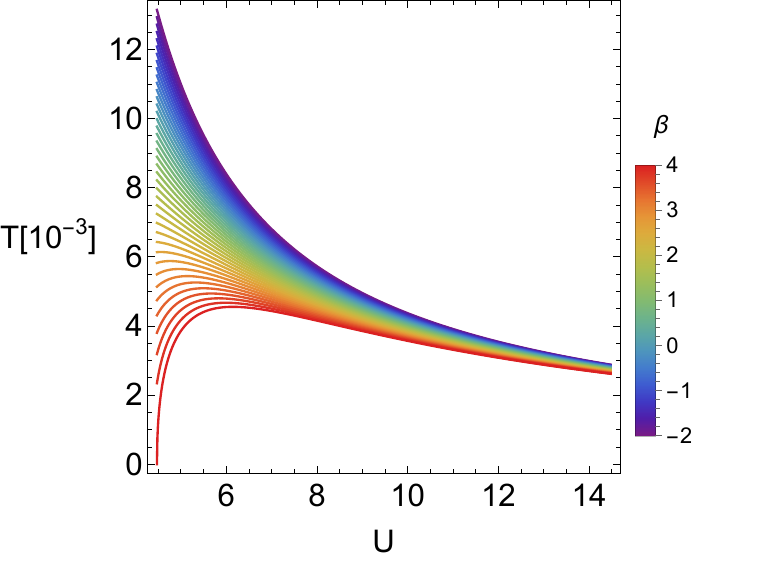} 
    (a)
\end{minipage}%
\hfill
\begin{minipage}[t]{0.5\linewidth}
    \centering
    \includegraphics[width=1.1\textwidth]
    {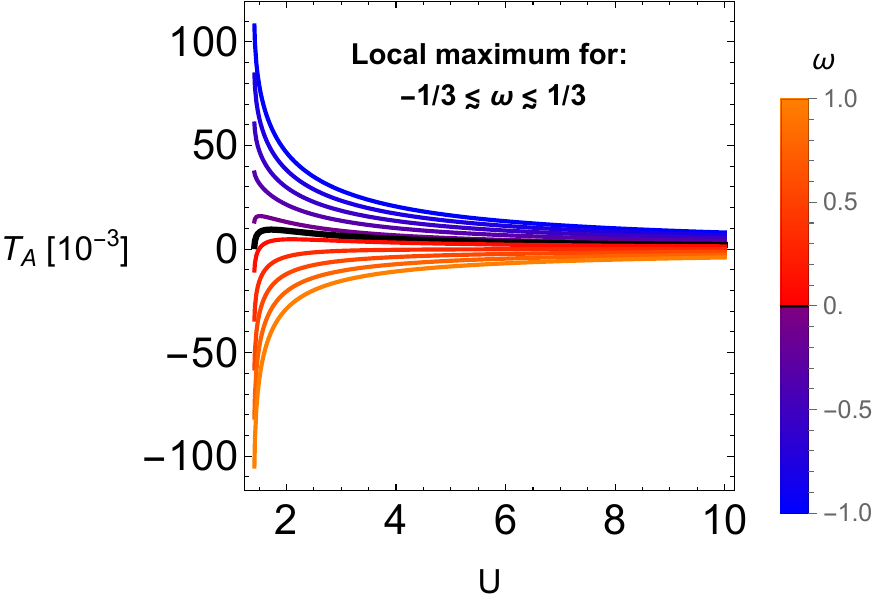} 
    (b)
\end{minipage}%
\hfill
\caption{(a) Thermodynamic temperature profile for different values of $\beta$. (b) Geometric temperature profile for $\beta>0$ and different values of $\omega$.}
\label{temp}
\end{figure}

As shown in Fig.~\ref{temp}(a), for \(\beta \leq 0\), the thermodynamic temperature  \eqref{temperature} is a smooth, monotonic function, allowing configurations to reach arbitrarily high temperatures while accessing correspondingly low values of internal energy and entropy. In contrast, for \(\beta > 0\), the temperature develops a local maximum, and the GUP parameter imposes a lower bound on the accessible energy and entropy of the system. Remarkably, the resulting temperature profile for \(\beta > 0\) exhibits a striking similarity to that observed in black hole models incorporating non-commutative geometry~\cite{nicolini2006noncommutative, modesto2010charged,araujo2024effects}. This suggests that, within gravitational systems, both the GUP and non-commutative geometry can induce comparable thermodynamic corrections, as originally pointed out in~\cite{nozari2007thermodynamics}. On the other hand, in Fig.~\ref{temp}(b) we illustrate the geometric temperature\footnote{The plot $T_A$ for $\beta\leq0$ is not presented since the local extrema only occur for negative temperatures.} $T_A$ for $\beta>0$. We observe that $T_A>0$ for $\omega \lesssim 1/3$ and $T_A$ has a local maximum for $-1/3 \lesssim \omega \lesssim 1/3$.

\begin{table}[H]
\centering
\renewcommand{\arraystretch}{2}
\begin{tabular}{|c||c||c||c|}
\hline
 & \makecell{\textbf{$T(U,\beta)$} \\ {(Eq.~\eqref{temperature})}}
 & \makecell{{$\widetilde{T}(U,V)$} \\ {(Eq.~\eqref{Work dens2})}}
 & \makecell{\textbf{$T_A(U,\beta,\omega)$} \\ {(Eq.~\eqref{geometric tempe})}} \\
\hline\hline

$\beta>0$
& local maximum in $U\approx 1.94 \sqrt{\beta}$
& no extrema
& local maximum only for $-1/3 \lesssim \omega \lesssim 1/3$ \\
\hline

$\beta=0$
& no extrema (GR)
& no extrema (GR)
& $T_A = 2T$ for $\omega=-1$ (GR) \\
\hline

$\beta<0$
& no extrema
& no extrema
& local minimum ($T_A<0$) \\
\hline
\end{tabular}
\caption{Comparison of temperature behavior for the three different equations of state (EOS) under the sign of \( \beta \).}
\label{tab:entropy_comparison}
\end{table}
Finally, Table~\ref{tab:entropy_comparison} summarizes the qualitative behavior of the three temperature functions associated with the different equations of state (EoS). As shown, the EOS defined by Eq.~\eqref{temperature} develops a local maximum only for \( \beta>0 \), while the alternative EoS \( \widetilde{T}(U,\beta,V) \) is monotonic for all values of \( \beta \). The geometric temperature \( T_A \) also exhibits non-monotonic behavior in a restricted interval of the parameter \( \omega \). In what follows, we focus exclusively on the EoS given by Eq.~\eqref{temperature}, derived from the fundamental equation \eqref{funda}, since it reproduces the thermodynamic structure expected in both non-commutative spacetimes and GUP–modified frameworks.

\subsection{Local thermodynamic stability}
In classical thermodynamics, local stability and the nature of phase transitions are determined by the behavior of the heat capacity and response functions~\cite{callen1998thermodynamics}. Heat capacities quantify the amount of energy exchanged with the surroundings during thermal processes and depend on the thermodynamic variables held fixed~\cite{romero2024extended1,Ladino:2024ned}. Stable configurations are typically associated with positive heat capacities, whereas negative values indicate thermodynamic instability. In the present case,  from the first law, Eq.~\eqref{first_law}, we observe that, within this thermodynamic space, the relevant heat capacity is defined at fixed \(\{\beta, \sigma\}\), and is given by

\begin{equation}
    C_{\beta \sigma}\equiv T\left(\frac{\partial S}{\partial T}\right)_{\beta\sigma}=-\frac{1}{T^2}\left(\frac{\partial^2 S}{\partial U^2}\right)^{-1}=\frac{4 \pi \sqrt{U^2-2\beta} \left[ \beta + U \left(U + \sqrt{U^2-2\beta}\right) \right]^2}
{ \beta \left(5U + 2\sqrt{U^2-2\beta}\right)-U^2 \left( \sqrt{U^2-2\beta}+ U\right)}.  
\end{equation}

\begin{figure}[H]
\begin{minipage}[t]{0.45\linewidth}
 \centering
\hspace{1cm}
\includegraphics[height=7cm]{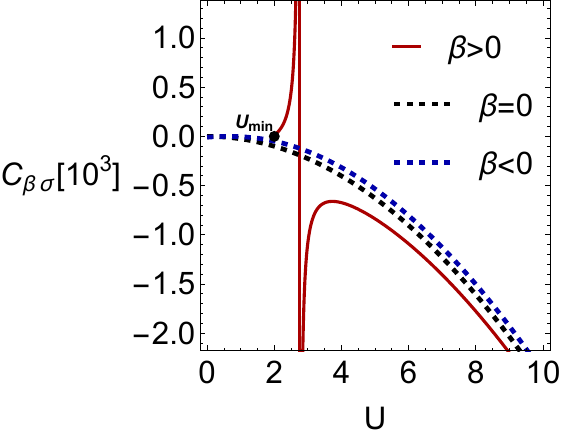}
 (a)\hspace{10cm}
\end{minipage}%
\hfill%
\begin{minipage}[t]{0.45\linewidth}
 \centering
\hspace{1cm}
\includegraphics[height=7cm]
{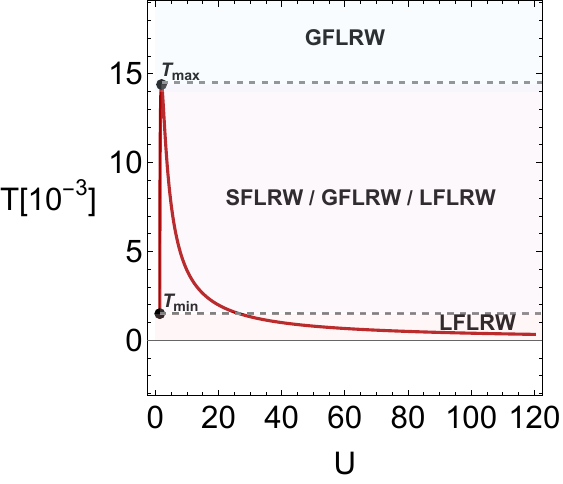}
(b)\hspace{12cm}
\end{minipage}%
\hfill%

  \caption{Heat capacity for different values of GUP parameter: \( \beta = 2 \) (red curve), $\beta=0$ (black curve), and \( \beta = -1.5 \) (blue curve). (b) Phase structure for \( \beta > 0 \). The horizon exhibits distinct thermodynamic regimes: SFLRW, LFLRW, and GFLRW. 
 } 
 \label{heat}
\end{figure}

\begin{figure}[H]
  \centering
 
  \begin{minipage}[t]{0.7\linewidth}
    \centering
    \includegraphics[height=7.5cm]
    {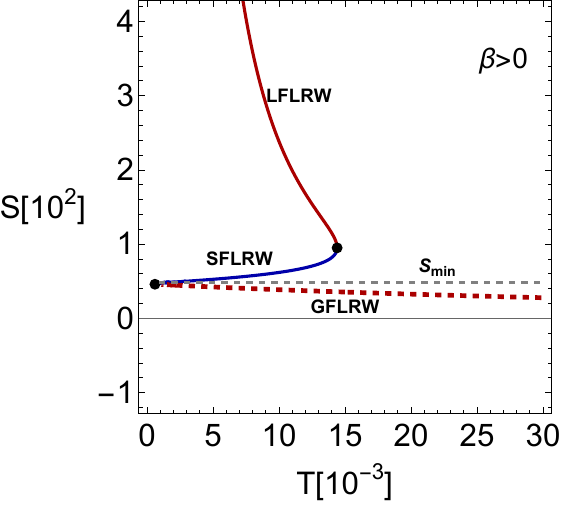}
    
  \end{minipage}
 \caption{ $S$--$T$ plot for $\beta = 1$ and $\sigma = 1$. The dashed red line represents the GFLRW phase.}
\label{fig:branches entropy}
\end{figure}

As illustrated in Fig.~\ref{heat}(a), for \( \beta > 0 \), the heat capacity is defined only for \( U \geq \sqrt{2\beta} \) and displays a single divergence at the local maximum of the temperature, located at \( U \approx 1.94\,\sqrt{\beta} \). This divergence signals a change in the local thermodynamic stability and suggests a possible phase transition in the horizon. The behavior implies that the deformation parameter \(\beta\) serves as a driving mechanism for the emergence of thermodynamic phase transitions. Conversely, for \(\beta \leq 0\), the heat capacity \( C_{\beta \sigma} \) remains negative throughout, indicating an unstable configuration. \\\\
This thermodynamic behavior resembles that of a vdW fluid, as depicted in Fig.~\ref{heat}(b), displaying two distinct and physically admissible thermodynamic phases for \( \beta > 0 \). In this analogy, the small-FLRW (SFLRW) phase, characterized by lower entropy, corresponds to a liquid-like state, whereas the large-FLRW (LFLRW) phase, associated with higher entropy, mirrors the behavior of a gaseous phase. For temperatures above \( T_{\max} \), the system enters a regime where \( 0 < S < S_{\min} \), which we refer to as the GUP-forbidden FLRW phase (GFLRW). This unphysical region arises directly from the lower bound on entropy imposed by the GUP. Although this phase does not manifest explicitly in the \( T \)--\( U \) diagram, it becomes evident in the \( S \)--\( T \) plane, as shown in Fig.~\ref{fig:branches entropy}. In contrast, for \( \beta \leq 0 \), the temperature becomes a monotonic function of the internal energy, and the distinction between the small and large phases disappears. The system then transitions into a regime analogous to a single-phase fluid, similar to those observed in condensed matter systems at supercritical temperatures~\cite{callen1998thermodynamics}. \\\\
While this analogy is insightful, it is important to note that, unlike the vdW fluid, the quasi-homogeneous GUP-FLRW system does not exhibit critical behavior in the \( T \)--\( U \), \( T \)--\( \beta \), or \( T \)--\( S \) thermodynamic planes. However, as discussed in~\cite{feng2024phase}, such critical phenomena instead emerge within the \( P \)--\( V \) space.

\subsection{Free energy and global thermodynamic stability}
\label{seccion_global}
\begin{figure}[H]
  \centering
  \begin{minipage}[t]{0.45\linewidth}
    \centering
    \includegraphics[height=7cm]
    {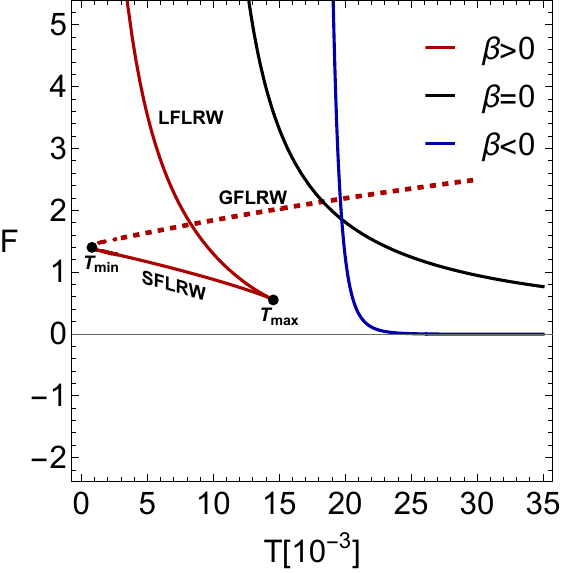}
    \caption*{(a)}
  \end{minipage}%
  \hfill
  \begin{minipage}[t]{0.45\linewidth}
    \centering
    \includegraphics[height=7cm]
    {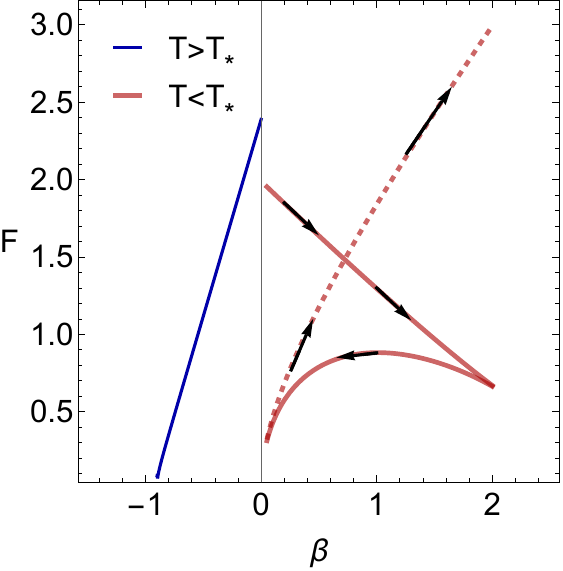}
    \caption*{(b)}
  \end{minipage}%
  \hfill\\
 \caption{Free energy for $\sigma = 1$. (a) $F$--$T$ plot for $\beta = 1$ (red), $\beta = 0$ (black), and $\beta = -1$ (blue), showing three branches for $\beta > 0$ and a single branch for $\beta \leq 0$. (b) $F$--$\beta$ isotherms for $T = 0.26$ (blue) and $T = 0.01$ (red); arrows indicate decreasing entropy, with three coexisting phases below $T_{\star}$ and one above. Curves are rescaled for clarity.}
\label{fig:free energy}
\end{figure}

A more comprehensive understanding of the global phase structure can be achieved by analyzing the Helmholtz free energy, defined via the standard thermodynamic relation \( F \equiv U - T S \)
\begin{equation}
F(T, \beta, \sigma) = \frac{8 \pi \beta + S + 2 K \left( 12 \pi \beta - S + 8 \pi \beta K \right)}{16 \pi \sqrt{\beta K} \left( 1 + K \right)}.
\end{equation}
The condition \( K < 1/2 \) defines the GFLRW branch, characterized by \( 0 < S < S_{\min} \). By combining Eqs.~(\ref{temperature}) and~\eqref{funda}, one obtains an explicit analytic expression for the free energy as a function of temperature. Figure \ref{fig:free energy} displays the full phase structure associated with the cosmological horizon. For \( \beta > 0 \), the behavior of the free energy closely mirrors that found in black holes within the extended phase space of AdS solutions~\cite{ Ladino:2024ned, LADINO2025117031}, as well as in anisotropic extensions of AdS spacetimes, such as hyperscaling violating black holes~\cite{herrera2021hyperscaling,herrera2023anisotropic}. In both scenarios, a metastable branch emerges from the oscillatory region of the equation of state, signaling the presence of a first-order phase transition.

In Fig.~\ref{fig:free energy}(a), we observe that for temperatures below \( T_{\min} \), only the LFLRW phase exists, which is thermodynamically unstable due to its negative heat capacity. Within the intermediate temperature range \( T \in [T_{\min}, T_{\max}] \), the three phases SFLRW/GFLRW/LFLRW can, in principle, coexist, forming an equilibrium mixture analogous to the liquid and vapor phases in conventional fluids. The SFLRW phase, being thermodynamically stable (positive heat capacity), is globally preferred as it minimizes the free energy. For temperatures above \( T_{\max} \), only the GFLRW branch remains, although it represents a non-physical, mathematically allowed solution.\\\\
 In summary, the \( F \)-\( T \) diagram displays the characteristic swallowtail structure, where the free energy curve intersects itself, an unmistakable signature of a first-order phase transition between the SFLRW and LFLRW phases, reminiscent of the gas–liquid transition in vdW fluids~\cite{callen1998thermodynamics}. In Fig.~\ref{fig:free energy}(b), we display the free energy isotherms. Although a closed-form expression for the temperature $T_{\star}$, above which the horizon behaves as a single fluid phase, could not be obtained, the key observation is that this single-phase regime occurs only for $\beta \leq 0$.
\begin{figure}[H]
  \centering
  \begin{minipage}[t]{0.45\linewidth}
    \centering
    \includegraphics[width=\linewidth]{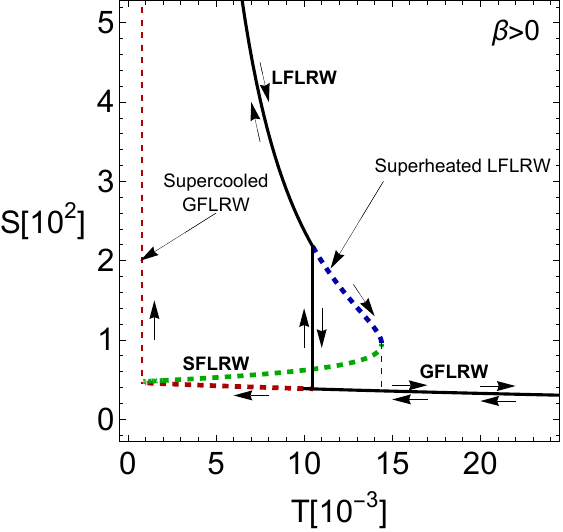}
    \caption*{(a)}
  \end{minipage}%
  \hfill
  \begin{minipage}[t]{0.4\linewidth}
    \centering
    \includegraphics[width=\linewidth]{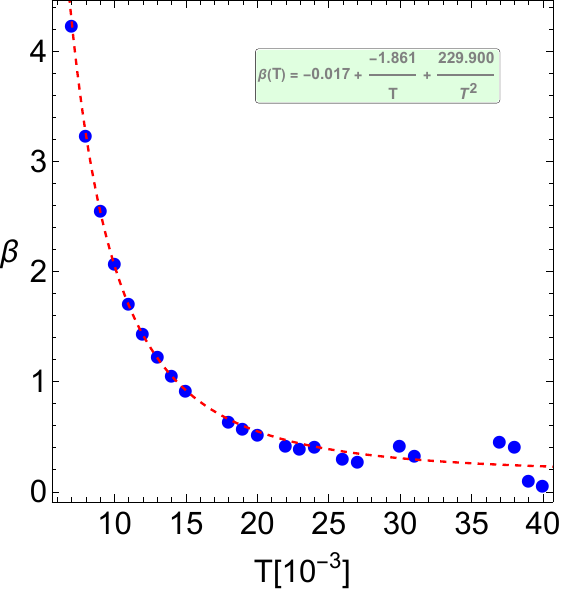}
    \caption*{(b)}
  \end{minipage}
 \caption{(a) The equilibrium behavior of $S$--$T$ plane is represented by the solid black curve. The vertical black line delineates the coexistence temperature. The dashed blue curve corresponds to the superheated LFLRW phase, the dashed green curve to the SFLRW phase, and the dashed red curve to the supercooled GFLRW phase. (b) Numerical solution of the coexistence curve. In both panels, $\sigma = 1$. }
\label{phases}
\end{figure}
It is well established that systems undergoing first-order phase transitions may exhibit thermal hysteresis, wherein the transition temperature, and consequently other thermodynamic properties, depends on the direction of the process. Specifically, during cooling, the transition can occur at a temperature below the equilibrium value (supercooling), while during heating, it may occur at a higher temperature (superheating). The vdW fluid serves as a prototypical model that captures this behavior~\cite{langer1969statistical,callen1998thermodynamics}. By extending the interpretation of the cosmological horizon as an effective vdW fluid. In Fig.~\ref{phases}(a), the vertical dashed blue and red lines denote irreversible, non-equilibrium transitions: from the superheated LFLRW phase to the GFLRW phase, and from the supercooled GFLRW phase to the LFLRW phase, respectively. The coexistence curve, along which both thermodynamic phases share the same free energy and temperature, cannot be obtained analytically. Therefore, in Fig.~\ref{phases}(b), we present a numerical solution for the coexistence curve, together with a fitting model of the form \( a_1 + a_2 T^{-1} + a_3 T^{-2} \), which accurately captures its behavior in the low-temperature regime.

\section{Geometrothermodynamics (GTD)}
\label{GTD}
Geometrothermodynamics is a formalism introduced in \cite{quevedo2007geometrothermodynamics} to geometrically investigate the thermodynamic properties of any physical system. A significant feature of GTD, absent in other geometric approaches such as Ruppeiner thermodynamics~\cite{ruppeiner1979thermodynamics,ruppeiner1981application}, is its invariance under Legendre transformations. In the context of ordinary thermodynamics, this invariance ensures that the physical properties of the system remain independent of the choice of thermodynamic potential used for its description \cite{callen1998thermodynamics}. Legendre invariance is incorporated into the formalism by introducing an auxiliary \((2n+1)\)-dimensional differential manifold   $\mathcal{T}$~\cite{quevedo2007geometrothermodynamics}. As a differentiable manifold, \(\mathcal{T}\) can be coordinatized by the set \(Z^A = \{ \Phi, E^a, I_a \}\), with \(a = 1, \ldots, n\), where \(n\) represents the number of thermodynamic degrees of freedom of the system, and \(\Phi\) denotes the thermodynamic potential. In the case of homogeneous systems, the variables \(E^a\) and \(I_a\) are clearly interpreted as extensive and intensive variables, respectively. However, for quasi-homogeneous systems, such as black holes or the GUP-FLRW universe, this distinction is less straightforward (see~\cite{quevedo2023unified,bravetti2017zeroth}). The space $\mathcal{T}$ is endowed with both a local canonical 1-form \(\Theta = d\Phi - I_a \, dE^a\), which satisfies the condition of being non-maximally integrable, defining a contact structure,
and a Riemannian metric $G=G_{AB}dZ^A d Z^B$ where $A,B=0,\ldots,2n$. The triplet \((\mathcal{T}, \Theta, G)\) defines a Riemannian contact manifold and is referred to as the thermodynamic phase space in the GTD formalism. Currently, there exist three Legendre-invariant metrics on \(\mathcal{T}\), which are given by

\begin{equation}
    G^{I/II}=\left(d\Phi-I_adE^a\right)^2+(\xi_{ab}E^aI^b)(\chi_{cd}dE^cdI^d), \label{metrics phase1}
\end{equation}
\begin{equation}
    G^{III}=\left(d\Phi-I_adE^a\right)^2+\sum_{a=1}^{n}\xi_{a}(E_aI_a)^{2k+1}dE^adI^a. \label{metric phase2}
\end{equation}
Here, \(\xi_a\) are \(n\) real constants, \(\xi_{ab}\) is a real diagonal \(n \times n\) matrix, and \(k\) is an integer. Moreover, the matrix \(\chi_{cd}\) is defined as \(\chi_{cd} = \delta_{cd} = \text{diag}(1, 1, \ldots, 1)\) for the metric \(G^I\), and as \(\chi_{cd} = \eta_{cd} = \text{diag}(-1, 1, \ldots, 1)\) for \(G^{II}\). In GTD, thermodynamic states are represented as points in an \( n \)-dimensional subspace of \( \mathcal{T} \), known as the equilibrium space \( \mathcal{E} \). This space is defined by a smooth mapping \( \varphi: \mathcal{E} \rightarrow \mathcal{T} \), for which the condition \( \varphi^*(\Theta) = 0 \) holds, where $\varphi^\ast$ represents the pullback. As a consequence, the first law of thermodynamics is naturally satisfied on \( \mathcal{E} \), and the coordinates \( Z^A \) become functions of the variables \( E^a \), that is $ Z^A(E^a) = \{ \Phi(E^a), E^a, I_a(E^a) \}$, where \( \Phi = \Phi(E^a) \)  represents the fundamental equation of the thermodynamic system~\cite{callen1998thermodynamics} and $I_a=\partial \Phi/\partial E^a $ the dual variables. Additionally, the line element $G = G_{AB}dZ^AdZ^B$ on $\cal{T}$ induces a line element $g = g_{ab}dE^adE^b$ on $\cal{E}$ by means of the pullback, i.e., $\varphi^\star(G) = g$. Then, from Eqs. (\ref{metrics phase1}) -- \eqref{metric phase2}, we obtain
\begin{align}
g^I &= \sum_{a,b,c=1}^{n} \left( \nu_c E^c \frac{\partial \Phi}{\partial E^c} \right) \frac{\partial^2 \Phi}{\partial E^a \partial E^b} \, dE^a \, dE^b,\label{g111} \\
g^{II} &= \sum_{a,b,c,d=1}^{n} \left( \nu_c E^c \frac{\partial \Phi}{\partial E^c} \right) \eta^d_{\;a} \frac{\partial^2 \Phi}{\partial E^b \partial E^d} \, dE^a \, dE^b, \label{g222}\\
g^{III}&=\sum_{a=1}^{n}\nu_a\left(\delta_{ad}E^d\frac{\partial \Phi}{\partial E^a}\right)^{2k+1}\delta^{ab}\frac{\partial^2 \Phi}{\partial E^b \partial E^c} dE^a dE^c. \label{g333}
\end{align}

As we are interested in describing quasi-homogeneous thermodynamic systems, we compute the components of the metrics on \( \mathcal{E} \) by choosing the free constants as \( \xi_a = \nu_a \) and \( \xi_{ab} = \text{diag}(\nu_1, \nu_2, \ldots, \nu_n) \), where \( \nu_a \) are the quasi-homogeneity coefficients that define the scaling properties of the fundamental equation. Furthermore, if the Euler relation, \( \Sigma_a \nu_a E^a \partial \Phi / \partial E^a = \nu_\Phi \Phi \), is satisfied, the conformal factor in $g^I$ and $g^{II}$ is replaced by \( \nu_\Phi \Phi \), where \( \nu_\Phi \) is the quasi-homogeneity degree of the potential \( \Phi \)~\cite{quevedo2023unified}. Additionally, to ensure that the three metrics describe the same thermodynamic system, we fix $k=0$ in $g^{III}$. With these considerations, the metrics take the form

\begin{align}
    g^I_{ab}&=\nu_\Phi \Phi \delta^c_a\frac{\partial^2 \Phi}{\partial E^b \partial E^c}, \label{g1}\\
    g^{II}_{ab}&=\nu_\Phi \Phi \eta^c_a\frac{\partial^2 \Phi}{\partial E^b \partial E^c}. \label{g2}\\
g^{III}&=\sum_{a=1}^{n}\nu_a\left(\delta_{ad}E^d\frac{\partial \Phi}{\partial E^a}\right)\delta^{ab}\frac{\partial^2 \Phi}{\partial E^b \partial E^c} dE^a dE^c. \label{g3} 
\end{align}
Specifically, we interpret the curvature singularities of the  metrics Eqs.~(\ref{g1})--(\ref{g3}) as corresponding to phase transitions in the thermodynamic system. For the quasi-homogeneous metrics $g^I$ and $g^{II}$ with an arbitrary number of thermodynamic degrees of freedom, we can write the the Ricci scalar 
\begin{align}
 \mathcal{ R^I}(E^1,E^2,E^3,\ldots, E^n)\quad &\propto  \quad\begin{vmatrix}
 \Phi_{,11} &  \Phi_{,12} & \ldots&  \Phi_{,1n}\\
\Phi_{,12} & \Phi_{,22} & \ldots &\Phi_{,2n}\\
\vdots &\vdots& \vdots&\vdots\\
\Phi_{,1n}&\Phi_{,2n}&\ldots&\Phi_{,nn}
\end{vmatrix}	^{-2},\\
\mathcal{R^{II}}(E^1,E^2,E^3,\ldots, E^n)\quad &\propto  \quad \begin{vmatrix}
 \Phi_{,11} &  0 & \ldots&  0\\
0 & \Phi_{,22} & \ldots &\Phi_{,2n}\\
\vdots &\vdots& \vdots&\vdots\\
0&\Phi_{,2n}&\ldots&\Phi_{,nn}
\end{vmatrix}	^{-2}.
\end{align}
The general structure of these scalars has been analyzed in \cite{Ladino:2024ned,LADINO2025117031}, where we obtained the general conditions that relate the singularities of the three GTD metrics\footnote{Notation: $\Phi_{,xy}\equiv \partial^2 \Phi/\partial x \partial y$.}, which for a three-dimensional thermodynamic space can be written as 
  \begin{align}
      I&:\Phi_{,11}\Big [ \big (\Phi_{,23}\big)^2-\Phi_{,22}\Phi_{,33} \Big]+\Phi_{,22}\big (\Phi_{,13}\big)^2+\Phi_{,33}\big (\Phi_{,12}\big)^2 -2 \Phi_{,12} \Phi_{,13} \Phi_{,23}=0,\label{c11}\\
       II&:\Phi_{,11}\Big [ \big (\Phi_{,23}\big)^2-\Phi_{,22}\Phi_{,33} \Big]=0,\label{vc}\\
      III:&\;\left\{
\begin{array}{l}
\Phi_{,11}=\Phi_{,12}= \Phi_{,13}=0, \\[6pt]
\Phi_{,12}= \Phi_{,22}=0, \\[6pt]
\Phi_{,13}= \Phi_{,33}=0, \\[6pt]
\Phi_{,22}=\Phi_{,33}=0.
\end{array}
\right.\label{c33}\
 \end{align}

The simultaneous fulfillment of the above conditions ensures that the curvature singularities of all three GTD metrics are mutually consistent \cite{quevedo2023unified}. Moreover, we can relate the singularities of the scalar curvature with the divergences of the response functions of the thermodynamic system. In ordinary thermodynamics, the response functions define second-order phase transitions and are essentially determined by the behavior of the independent variables $E^a$ in terms of their duals $I_a$. Thus, the singularity conditions $I$ (\ref{c11}), $II$ (\ref{vc}), and $III$ (\ref{c33}) in the energy representation take the following form (see Appendix~B in Ref.~\cite{Ladino:2024ned})
\begin{align}
I &: \quad \frac{T}{C_{I_2,I_3} \times \kappa_{S,I_3} \times \kappa_{S,E^2}} = 0, \label{re}\\[8pt]
II &: \quad \frac{T}{C_{E_2,E_3} \times \kappa_{S,I_3} \times \kappa_{S,E^2}} = 0, \\[15pt]
III &: \quad 
\left\{
\begin{array}{ll}
\text{(a)} & \dfrac{1}{C_{E^2,E^3}} = \dfrac{1}{\alpha_{S,E^2}} = \dfrac{1}{\alpha_{S,E^3}} = 0, \\[15pt]
\text{(b)} & \dfrac{1}{\alpha_{S,E^3}} = \dfrac{1}{\kappa_{S,I_3}} = 0, \\[15pt]
\text{(c)} & \dfrac{1}{\alpha_{S,E^2}} = \dfrac{1}{\kappa_{S,E^2}} = 0, \\[15pt]
\text{(d)} & \dfrac{1}{\kappa_{S,I_3}} = \dfrac{1}{\kappa_{S,E^2}} = 0.
\end{array}
\right.
\end{align}
where $C_{x_1 x_2}$, $\kappa_{x_1 x_2}$, and $\alpha_{xy}$ denote the heat capacity, the compressibility parameter, and the coefficient of thermal expansion, respectively, evaluated at a fixed set of thermodynamic parameters $\{x_1, x_2\}$, as defined in \cite{Ladino:2024ned}. Therefore, we observe that, in general, the singularities of the equilibrium space are associated with the phase transition structure of the system. 
In the forthcoming subsection, we  explore the GTD of the FLRW universe in pure Einstein gravity, and a cosmology modified by a GUP parameter.

\subsection{GTD of the FLRW universe in Einstein Gravity}
\label{perfect fluid}
In pure Einstein gravity, the fundamental equation in the energy representation is given by \cite{sanchez2023thermodynamics}
\begin{equation}
    U(S,V)=\frac{3}{8}\frac{V}{S}.\label{funda_fluid}
\end{equation}
Observe that for \( \beta = 0 \), the fundamental equation  \eqref{funda} reduces to Eq. \eqref{funda_fluid}, once we identify the thermodynamic volume with \( V = 4\pi R^3/3 \). The temperature and pressure can be directly obtained as 
\begin{equation}
 T(S,V)\equiv-\frac{\partial U(S,V)}{\partial S}=\frac{3V}{8S^2}, \quad  p(S, V) \equiv -\frac{\partial U(S,V)}{\partial V} = -\frac{3}{8S}.
\end{equation}
To study the GTD of the FLRW horizon we use \( \{S, V\} \) as independent fluctuation variables, for which Eq. \eqref{funda_fluid}, in general, is a quasi-homogeneous function of arbitrary degree \( \nu_U = \nu_V - \nu_S \). Since the Euler relation $\nu_S S U_{,S}+\nu_V V U_{,V}=\nu_U U$ is fulfilled, the GTD metrics given by Eqs.~(\ref{g1})--(\ref{g3}) read
\begin{align}
    g^I&= \nu_U U\Big(U_{,SS}
dS^2+2U_{,SV}dSdV
+U_{,VV}dV^2\Big),\label{metric21}\\ 
    g^{II}&= \nu_U U\Big(-U_{,SS}
dS^2+U_{,VV}dV^2\Big),\label{metric22}\\ 
    g^{III}&=\Big(\nu_S  S U_{,S} \big)U_{,SS}dS^2+\Big(\nu_U U\Big)U_{,SV}dSdV+\Big(\nu_V V U_{,V}\Big)U_{,V V}dV^2.
\end{align}
Using Eq.~\eqref{funda_fluid}, one finds that the Ricci scalar \( \mathcal{R} \) for the metrics \( g^I \) and \( g^{III} \) vanishes, showing that they describe flat thermodynamic equilibrium spaces. In contrast, the metric \( g^{II} \) becomes degenerate because the fundamental equation is linear in the volume, implying \( U_{,VV} = 0 \). Since the pressure is volume-independent, the isentropic compressibility diverges, \( \kappa_S \to \infty \), leading to the degeneracy of \( g^{II} \). This indicates that \( S \) and \( V \) are not suitable as independent variables. A similar situation occurs in Ruppeiner geometry for charged AdS black holes, where \( C_V = 0 \)~\cite{dehyadegari2020microstructure}. In both cases, the degeneracy obstructs a direct GTD analysis. To overcome this, following~\cite{wei2019repulsive,wei2019ruppeiner}, we rewrite the metric \( g^{II} \) as
\begin{equation}
 g^{II}= \nu_U U(S,V)\left(-\frac{3 V}{4S^3}
dS^2+\frac{1}{V\kappa_S} dV^2\right),\label{metric_normalized}
\end{equation}
where we have used 
\begin{equation}
    U_{,SS}=\frac{3V}{4S^3}, \quad \kappa_S\equiv -\frac{1}{V}\left(\frac{\partial V}{\partial P}\right)_{S}= \frac{1}{V U_{,VV} }.
\end{equation}

 Assuming \(\kappa_S\) to be constant, we probe the GTD of the horizon through the metric \( g^{II} \). To this end, we compute the scalar curvature from Eq.~\eqref{metric_normalized} and, using Eq.~\eqref{funda_fluid}, obtain
\begin{equation}
    \mathcal{R}^{II}(S,V) = -\frac{1}{4 \nu_U U(S,V)^2}.
\end{equation}
Since, for any physically acceptable configuration, we always have \(\nu_U U(S,V) \neq 0\). Consequently, \(\mathcal{R}^{II}\) remains regular, leading to an equilibrium manifold free of singularities. This implies that the metric \(g^{II}\) describes the apparent horizon of the FLRW universe as an interacting thermodynamic system without any phase transitions, which is the expected behavior in cosmology based on pure Einstein gravity, as discussed in~\cite{sanchez2023thermodynamics,debnath2020thermodynamics,cai2005first}. In the following section, we demonstrate that when thermodynamic fluctuations in $\beta$ are taken into account, the GTD predicts the occurrence of phase transitions.

\subsection{GTD of the GUP-FLRW universe}

\label{GUP GTD section}
For the GUP-FLRW system, the most general scenario involves considering a three-dimensional thermodynamic equilibrium space $\{U,\beta,\sigma\}$. While it is possible to work with a reduced two-dimensional equilibrium space, such an approach may yield inconsistent results, as it neglects fluctuations in specific thermodynamic parameters (see, for instance, \cite{mirza2007ruppeiner}). As, shown in Sec. \ref{funda_sect}, the fundamental equation Eq. \eqref{funda} is a quasi-homogeneous function. Thus, according to Eqs.~(\ref{g1}) - (\ref{g3}), the line elements of the GTD metrics in the entropy representation can be written as 
\begin{align}
    g^I&= \nu_S S\Big(S_{,UU}
dU^2+2S_{,U\beta}dUd\beta+2S_{,U\sigma}dSd\sigma+2S_{,\beta\sigma}d\beta d\sigma+S_{,\beta\beta}d\beta^2
+ S_{,\sigma\sigma}d\sigma^2\Big),\\
 g^{II} &= \nu_S S\Big(-S_{,UU}
 dU^2+2S_{,\beta\sigma}d\beta d\sigma+S_{,\beta\beta}d\beta^2
+ S_{,\sigma\sigma}d\sigma^2\Big),\label{metric2}\\ 
    g^{III}&=\nu_U \big(U/T \big)S_{,UU}dU^2+\nu_\beta \big(B\beta/T \big)S_{,\beta \beta}d\beta^2 +\nu_\sigma \big(\Sigma \sigma/T \big)S_{,\sigma \sigma}d\sigma^2\notag\\&+S_{,U\beta}\Big[\nu_U \big(U/T \big)+\nu_\beta \big(B\beta/T   \big)\Big]dUd\beta +S_{,U\sigma}\Big[\nu_U \big(U/T \big)\\&+\nu_\sigma\big(\Sigma \sigma/T\big)\Big]dUd\sigma+S_{,\beta \sigma}\Big[\nu_\beta \big(B\beta/T  \big)+\nu_\sigma \big(\Sigma \sigma/T\big)\Big]d\beta d\sigma.\notag 
\end{align}
 The singularity conditions Eqs. \eqref{c11} - \eqref{c33} are written as
\begin{align}
I:&\; S_{,UU}\Big[ \big( S_{,\beta \sigma} \big)^2 - S_{,\beta \beta} S_{,\sigma \sigma} \Big]
    + S_{,\beta \beta} \big( S_{,U\sigma} \big)^2
    + S_{,\sigma \sigma} \big( S_{,U\beta} \big)^2
    - 2 S_{,U\beta} S_{,U\sigma} S_{,\beta \sigma} = 0, \label{RNcondI} \\[6pt]
II:&\; S_{,UU}\Big[ \big( S_{,\beta \sigma} \big)^2 - S_{,\beta \beta} S_{,\sigma \sigma} \Big] = 0, \\[6pt]
III:&\;
\left\{
\begin{array}{ll}
\text{(a)} & S_{,UU} = S_{,U\beta} = S_{,U\sigma} = 0, \\[6pt]
\text{(b)} & S_{,U\beta} = S_{,\beta\beta} = 0, \\[6pt]
\text{(c)} & S_{,U\sigma} = S_{,\sigma\sigma} = 0, \\[6pt]
\text{(d)} & S_{,\beta\beta} = S_{,\sigma\sigma} = 0.
\end{array}
\right.
\label{RNcondIII}
\end{align}

\begin{figure}[H]
\begin{minipage}[t]{0.32\linewidth}
 \centering
\hspace{1cm}
\includegraphics[width=1\linewidth]{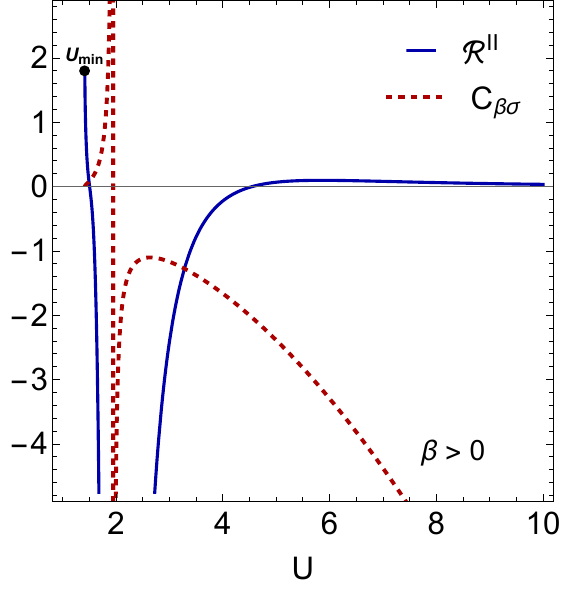}
 (a)\hspace{10cm}
\end{minipage}%
\hfill%
\begin{minipage}[t]{0.32\linewidth}
 \centering
\hspace{1cm}
\includegraphics[width=1.05\linewidth]{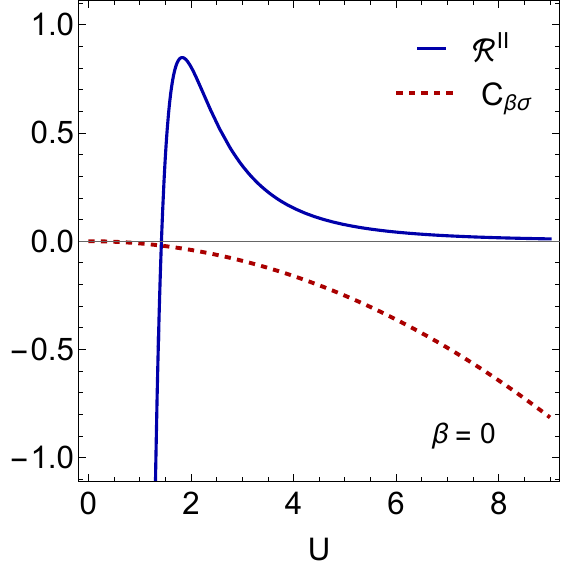}
(b)\hspace{12cm}
\end{minipage}%
\hfill%
\centering
\begin{minipage}[t]{0.32\linewidth}
 \centering
\hspace{1cm}
\includegraphics[width=1\linewidth]{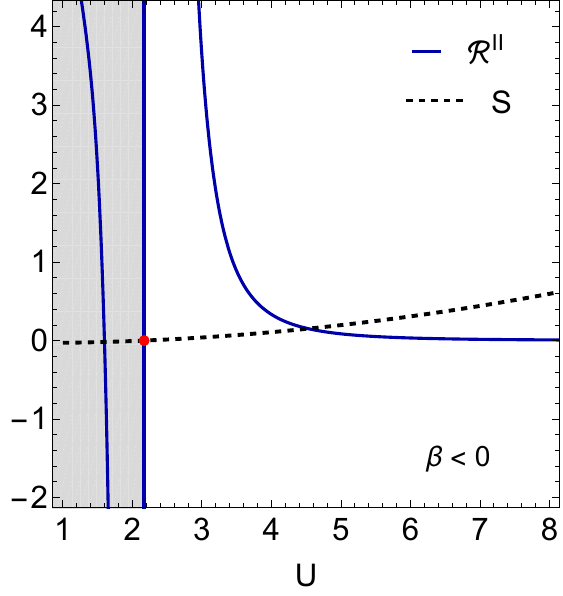}
(c)\hspace{12cm}
    \end{minipage}%
\hfill%
\caption{$\mathcal{R}^{II}(U,\beta,\sigma)$ and heat capacity $C_{\beta\sigma}$ as functions of $U$ for different GUP parameters, with $\sigma=1$ and $\beta_S=1$ fixed. Panels correspond to: (a) $\beta = 1$, (b) $\beta = 0$, and (c) $\beta = -1$. The gray region indicates $S<0$. The plots have been rescaled for clarity.}
\label{RII}
\end{figure}

Using Eq. \eqref{funda} and  shorthand notation $X \equiv \sqrt{U^{2} - 2\beta}$, $Y\equiv U+X$, the conditions read 
\begin{align}
I:&\;  \frac{32 \pi^3 (\beta - 2 U^2) \left( \beta + U Y \right)}{\sigma^2 X^4} = 0, \\[6pt]
II:&\; \frac{32 \pi^3 U (\beta - 2 U^2) \left[ YU^2  - \beta \left( 5 U + 2 X \right) \right]}{\sigma^2 X^3} = 0, \\[6pt]
III:&\;
\left\{
\begin{array}{l}
\text{(a)}\quad 2XY^{2} - U\bigl(Y^2 + 4\beta\bigr)
= -2XY + 4X^2 + Y^2 + 4\beta = 0, \\[6pt]
\text{(b)}\quad -2XY + 4X^2 + Y^2 + 4\beta
= \dfrac{2\pi}{X^2} - \dfrac{2\pi Y}{X^3} - \dfrac{16\pi}{XY}
- \dfrac{8\pi\beta}{X^2Y^2} - \dfrac{8\pi \beta}{YX^2}=0, \\[8pt]
\text{(c)}\quad \dfrac{4\pi \beta}{\sigma^2}=0, \\[6pt]
\text{(d)}\quad \dfrac{2\pi}{X^2} - \dfrac{2\pi Y}{X^3} - \dfrac{16\pi}{XY}=0.
\end{array}
\right.
\end{align}
\noindent
Condition~$I$, as well as all components of Condition~$III$, admit no real solution consistent with \(U>0\) and \(\beta \neq 0\). Consequently, both \(g^{I}\) and \(g^{III}\) describe regular thermodynamic manifolds. In contrast, in $II$, the condition $S_{,UU}=YU^2  - \beta \left( 5 U + 2 X\right)=0$ coincides with the critical curve of the heat capacity $C_{\beta \sigma}$, with the exact solution $U \approx 1.94\sqrt{\beta}$. Moreover, in $II$, the condition $ \left( S_{,\beta \sigma} \right)^2 - S_{,\beta \beta} S_{,\sigma \sigma}=U (-\beta + 2 U^2)=0$ cannot be satisfied within the physical domain $X>0$, thereby ensuring the absence of additional singularities in $g^{II}$.
For the remainder of this work, we restrict our analysis to the Ricci scalar $\mathcal{R}^{II}$, as it provides a precise characterization of the phase structure in the thermodynamic space $\{U, \beta, \sigma\}$, and is explicitly given by
\begin{equation}
    \mathcal{R^{II}}(U,\beta,\sigma)=\frac{\mathcal{N^{II}}}{\mathcal{D^{II}}},
\end{equation}
where the numerator takes the form
\begin{equation}
\begin{aligned}
\mathcal{N}^{II} =&\, 4\beta^9 - 1184\, U^{17} Y - 2\beta^8 U (352U + 37 X) \\
&+ 32\beta\, U^{15} (395U + 358 X) + 72\beta^3 U^{11} (1883U + 1317 X) \\
&+ \beta^7 U^3 (8943U + 1993 X) - 16\beta^2 U^{13} (3547U + 2868 X) \\
&+ 14\beta^5 U^7 (9453U + 4196 X) - \beta^6 U^5 (49785U + 15934 X) \\
&- 2\beta^4 U^9 (90529U + 52095 X) \\
&- 4X^2 \ln\left( \frac{4\pi Y^2}{\sigma} \right) \Bigg\{ -2\beta^8 - 64 U^{15} Y+ 3\beta^7 U (81U + 11 X) \\
&+ 32\beta U^{13} (31U + 29 X) + 22\beta^5 U^5 (445U + 167 X) \\
&- 16\beta^2 U^{11} (327U + 271 X) - 2\beta^6 U^3 (1193U + 298 X) \\
&+ 8\beta^3 U^9 (1655U + 1167 X)- 4\beta^4 U^7 (4183U + 2285 X) \\
&+ \beta \ln\left( \frac{4\pi Y^2}{\sigma} \right) \Bigg[ 2\beta^7 - 16U^{13}Y + 16\beta U^{11} (23U + 22 X) \\
&- \beta^6 U (183U + 29 X) - 8\beta^2 U^9 (263U + 220 X) \\
&+ \beta^5 U^3 (1415U + 391 X) + 4\beta^3 U^7 (1157U + 759 X) \\
&- 2\beta^4 U^5 (2095U + 934 X)\Bigg] \Bigg\};
\end{aligned}
\end{equation}
and the denominator is given by
\begin{equation}
\begin{aligned}
\mathcal{D}^{II} = \frac{1}{\pi} \, \nu_S \,  U^2 S(U, \beta, \sigma)^3 
\left(  \beta-U Y  \right)^2 \, 
\left( \beta - 2U^2 \right)^2 
\big[ YU^2 - \beta (5U + 2X) \big]^2.
\end{aligned}
\end{equation}
In Fig.~\ref{RII}, we present the results of the GTD analysis using $\mathcal{R^{II}}$. The GTD formalism captures the thermodynamic behavior of the FLRW universe with the GUP parameter \(\beta\), including divergences and possible phase transitions. For \(\beta < 0\), Fig.~\ref{RII}(c) shows that the scalar curvature \(\mathcal{R}^{II}\) remains regular, consistent with the heat capacity and indicating the absence of phase transitions. Although a divergence appears in this regime, it corresponds to vanishing entropy and thus lacks physical meaning. In contrast, Fig.~\ref{RII}(a) reveals that for \(\beta > 0\), a divergence in \(\mathcal{R}^{II}\) coincides with one in the heat capacity, confirming a genuine phase transition. For \(\beta = 0\), Fig.~\ref{RII}(b) shows that \(\mathcal{R}^{II}\) remains finite and nonzero, signaling thermodynamic interactions without phase transitions. In contrast, when fluctuations of \(\beta\) are neglected (Sec.~\ref{perfect fluid}), the metrics \(g^{I}\) and \(g^{III}\) yield flat manifolds while \(g^{II}\) degenerates, highlighting the necessity of including fluctuations in all thermodynamic degrees of freedom for a consistent GTD description.
\begin{table}[H]
\centering
\begin{tabular}{|c|c|c||c|c||c|c|}
\hline
GUP & \multicolumn{2}{c||}{Physical singularities} & \multicolumn{2}{c||}{Zeros} & \multicolumn{2}{c|}{Sign of $\mathcal{R^{II}}$ in physical region} \\
\hline
 & $2d$ & $3d$ & $2d$ & $3d$ & $2d$ & $3d$ \\
\hline
$\beta<0$ & 0  & 0 & 1 (entropy) & 1 (entropy) & - & + \\
$\beta=0$ & 0 & 0 & 0 & 1 & - & -/+ \\
$\beta>0$ &  1 (heat capacity) & 1 (heat capacity) & 1 & 2 & +/- & +/-/+ \\
\hline
\end{tabular}
\caption{Comparison between 2$d$ and 3$d$ GTD. Fluctuations of $\sigma$ are neglected in the two-dimensional case.}
\label{tab:GTD_comparison}
\end{table}
Finally, Table~\ref{tab:GTD_comparison} presents a comparative analysis, using the metric \(g^{II}\), between the two-dimensional (\(2d\)) and three-dimensional (\(3d\)) equilibrium spaces.
In the 2$d$ case, fluctuations of the integration constant $\sigma$ are neglected. In general, we observe that although $\sigma$ does not modify the divergences of the scalar curvature, and therefore does not alter the phase transition structure, it does affect the number of zeros of the scalar curvature. At present, these zeros lack a clear thermodynamic interpretation. However, as we discuss in the next section, this modification may affect the nature of the underlying microstructural interactions.

\section{Probing Thermodynamic Microstructure through GTD}
\label{micro}
\begin{figure}[H]
\begin{minipage}[t]{0.32\linewidth}
 \centering
\hspace{1cm}
\includegraphics[width=1\linewidth]{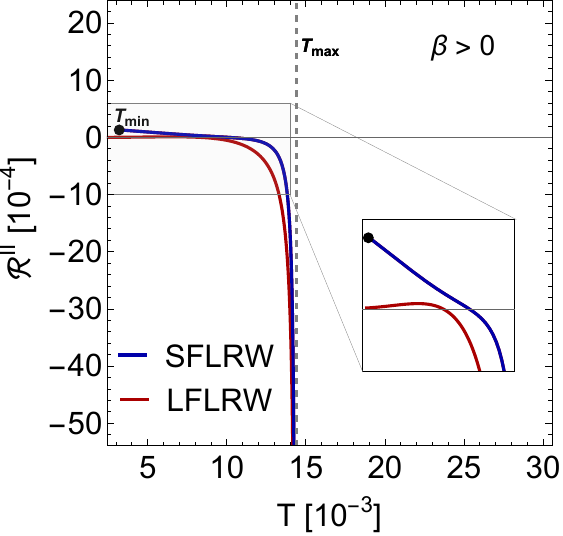}
 (a)\hspace{10cm}
\end{minipage}%
\hfill%
\begin{minipage}[t]{0.32\linewidth}
 \centering
\hspace{1cm}
\includegraphics[width=1\linewidth]{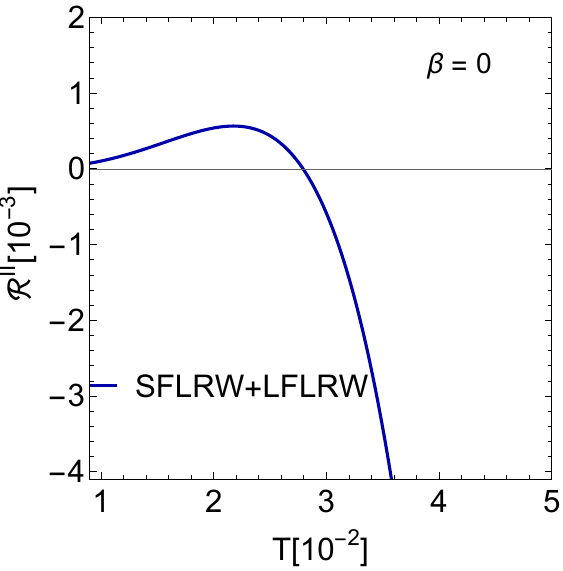}
(b)\hspace{12cm}
\end{minipage}%
\hfill%
\centering
\begin{minipage}[t]{0.32\linewidth}
 \centering
\hspace{1cm}
\includegraphics[width=1\linewidth,]{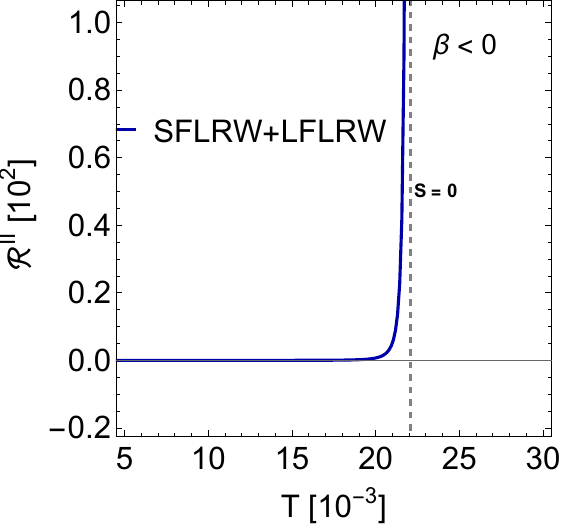}
(c)\hspace{12cm}
    \end{minipage}%
\hfill%
\caption{3$d$ GTD Scalar curvature $\mathcal{R}^{II}$ for different $\beta$, and $\sigma=1$, $\beta_S=1$. (a) For $\beta = 1$,  (b) for $\beta =0$, and (c) for $\beta = -1$.}
\label{microstructure}
\end{figure}

Due to the lack of a fully consistent statistical and quantum theory of gravity, alternative approaches have been proposed to gain insight into the microscopic behavior of gravitational systems. In this context, the scalar thermodynamic curvature has been introduced as a quantity related to the correlation length of a thermodynamic system~\cite{ruppeiner1995riemannian}, potentially capturing information about the underlying microscopic interactions.  In GTD, a positively curved equilibrium space ($\mathcal{R}^{II}>0$) is  associated with repulsive thermodynamic interactions, whereas a negative curvature ($\mathcal{R}^{II}<0$) indicates attractive thermodynamic interactions. In contrast, zero curvature ($\mathcal{R}^{II}=0$) corresponds to the absence of interactions, as in the case of an ideal gas, which is geometrically represented by a flat thermodynamic manifold~\cite{ruppeiner1979thermodynamics}. As shown in Fig.~\ref{microstructure}, the apparent horizon of the GUP-FLRW universe exhibits both attractive and repulsive thermodynamic interactions depending on the values of $T$ and $\beta$. In Fig.~\ref{microstructure}(a), we observe that for \(\beta > 0\), only the LFLRW branch exists at very low temperatures \(T < T_{\min}\), while in the intermediate range \(T \in [T_{\min}, T_{\max}]\), the scalar curvature becomes multivalued. In this regime, two distinct phases coexist, and a phase transition between the SFLRW and LFLRW branches is expected to occur at \(T_{\max}\). Conversely, in Fig.~\ref{microstructure}(b) and Fig.~\ref{microstructure}(c), we note that for \(\beta \leq 0\), the system enters an unstable fluid phase characterized by negative heat capacity (see, Fig. \ref{heat}). In this regime, no phase transition takes place since the thermodynamic phases are indistinguishable.
Additionally, it is observed that regardless of the sign of \(\beta\), thermodynamic interactions become stronger at higher temperatures. Remarkably, for \(\beta < 0\) (Fig.~\ref{microstructure}(c)), these interactions are always repulsive and are \(5\text{–}6\) orders of magnitude stronger than in the positive-\(\beta\) case. This amplification may be attributed to the fact that a negative GUP parameter enhances microscopic correlations by  allowing the system to access states with arbitrarily low entropy. Unlike the case \(\beta > 0\), which imposes a lower bound on entropy. However, further investigation is required to provide a more robust explanation. Compared with the Ruppeiner approach, the main difference in the GTD results is that, in Ruppeiner geometry, the interactions are always attractive for both phases, independently of the temperature \cite{feng2024phase}.
Similar discrepancies are also observed in the context of black holes, as discussed in \cite{LADINO2025117031,Ladino:2024ned}. Another noteworthy distinction, as we discuss next, lies in the critical exponent predicted by the thermodynamic curvature. Investigating the critical behavior of the GTD scalar curvature is notably more challenging than in Ruppeiner geometry, primarily because GTD metrics are not of Hessian form, and the thermodynamic potential appears explicitly in the metric components. In the GUP-FLRW model, analytical computation of temperature fluctuations using the free energy representation is not feasible. Nevertheless, a numerical analysis can be performed for both the LFLRW and SFLRW solutions in the vicinity of the maximum temperature, $T_{\text{max}}$. From this analysis, we extract the corresponding values of  $\mathcal{R}^{\mathrm{II}}$. Motivated by the behavior of correlation functions near criticality \cite{callen1998thermodynamics}, the resulting data are fitted to the phenomenological form
\begin{equation}
    \mathcal{R}^{\mathrm{II}}(T) \sim \frac{A}{(T - T_{\text{max}})^{\zeta}},
\end{equation}
where the fitted values of the amplitude \( A \) and the critical exponent \( \gamma \) are summarized in Table~\ref{tab:fit}.

\begin{table}[H]
\centering
\begin{tabular}{|c|c||cc||cc||cc||cc|}
\hline
$\beta$ & $T_{\text{max}}[10^{-3}]$ 
& \multicolumn{2}{c||}{2$d$ LFLRW} & \multicolumn{2}{c||}{3$d$ LFLRW} 
& \multicolumn{2}{c||}{2$d$ SFLRW} & \multicolumn{2}{c|}{3$d$ SFLRW} \\
\hline
 & & $A[10^{-6}]$ & $\zeta$ & $A[10^{-6}]$ & $\zeta$ 
 & $A[10^{-6}]$ & $\zeta$ & $A[10^{-6}]$ & $\zeta$ \\
\hline
1.0 & 14.40 & -1.01714 & 0.98756 & -1.28669 & 0.98814 & -0.64779 & 1.01844 & -0.78169 & 1.02223 \\
1.5 & 11.78 & -3.11097 & 0.78518 & -3.86769 & 0.78615 & -1.90096 & 0.81773 & -2.19107 & 0.82365 \\
2.0 & 10.20 & -1.49986 & 0.78219 & -1.84765 & 0.78322 & -0.86117 & 0.81886 & -0.95721 & 0.82661 \\
2.5 & 9.11  & -0.14227 & 0.94627 & -0.17581 & 0.94652 & -0.00742 & 0.99060 & -0.08009 & 1.00004 \\
3.0 & 8.31  & -0.37738 & 0.81224 & -0.45946 & 0.81329 & -0.19529 & 0.85610 & -0.20499 & 0.86693 \\
3.5 & 7.70  & -0.42872 & 0.76097 & -0.51662 & 0.76252 & -0.21505 & 0.80639 & -0.21973 & 0.81868 \\
4.0 & 7.20  & -0.00303 & 0.97395 & -0.03698 & 0.97407 & -0.00137 & 1.02839 & -0.01367 & 1.04206 \\
4.5 & 6.79  & -0.00581 & 0.88766 & -0.07013 & 0.88850 & -0.00263 & 0.94143 & -0.02574 & 0.95613 \\
5.0 & 6.44  & -0.00160 & 0.97832 & -0.01948 & 0.97850 & -0.00067 & 1.03784 & -0.00647 & 1.05389 \\
\hline
\end{tabular}
\caption{Fitted parameters for the 2$d$ and 3$d$ GTD scalar curvature $\mathcal{R}^{\mathrm{II}}$ near $T_{\text{max}}$, for various values of $\beta$ and $\sigma=1$.}
\label{tab:fit}
\end{table}

\begin{figure}[H]
  \centering
  \begin{minipage}[t]{0.45\linewidth}
    \centering
    \includegraphics[height=7cm]{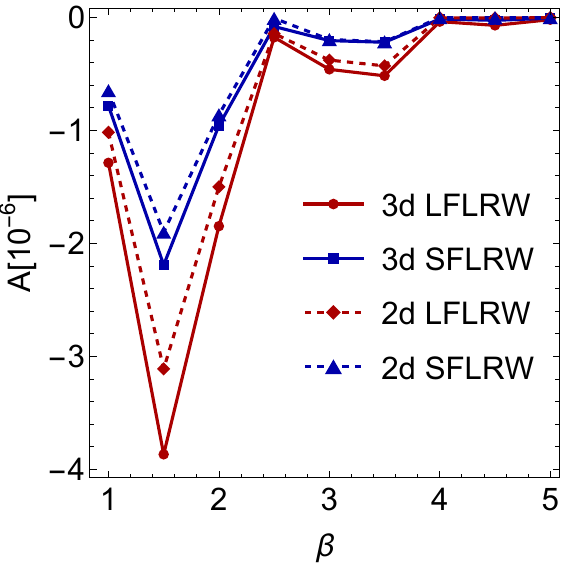}
    \caption*{(a)}
  \end{minipage}%
  \hfill
  \begin{minipage}[t]{0.45\linewidth}
    \centering
    \includegraphics[height=7cm]{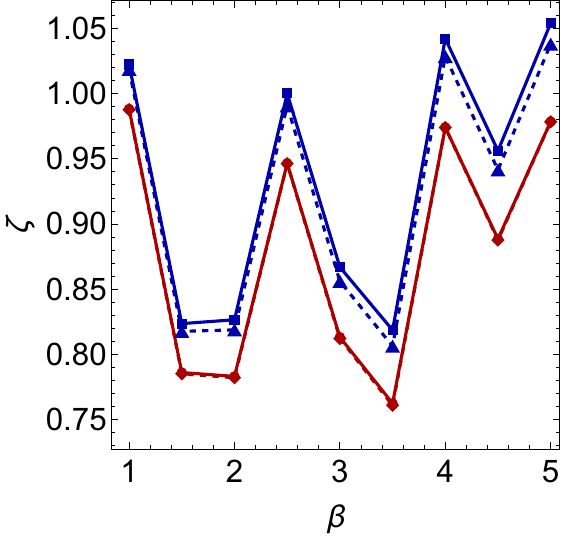}
    \caption*{(b)}
  \end{minipage}
\caption{Numerical results for the behavior of scalar $\mathcal{R}^{II}$ near $T_{\text{max}}$ for the 2$d$ and 3$d$ GTD cases.  
(a) Amplitude: As $\beta$ increases, the amplitudes for the LFLRW and SFLRW branches tend to converge.  
(b) Critical exponent $\zeta$: The values fluctuate around $\zeta \sim 1$, with a maximum deviation of approximately 25\%.}

\label{exponnt}
\end{figure}
From Fig.~\ref{exponnt} and Table~\ref{tab:fit}, we observe that the GTD critical exponent scales approximately as $\zeta \sim 1$, independently of the dimension of the equilibrium manifold. In contrast, Ruppeiner thermodynamic geometry yields exact values for the critical parameters: $A = -1/8$ and $\zeta = 2$, as discussed in \cite{feng2024phase}. Moreover, the results summarized in Table~\ref{tab:fit} indicate that varying $\beta$ may alter the exponent $\zeta$, potentially changing the underlying universality class of the system. A similar effect has been observed in the Ruppeiner geometry of higher-dimensional charged AdS black holes, where changes in the spacetime dimension modify the critical exponent \cite{wei2019ruppeiner}, as well as in the case of higher-dimensional singly spinning Kerr-AdS black holes \cite{wei2016analytical}.


\section{Conclusions}
\label{conc}
In this work, we developed a thermodynamic description of the FLRW universe incorporating quantum gravity effects through the specific GUP model introduced in~\cite{du2022new}. We observed that $P$–$V$ phase transitions can only occur for horizons with $\dot{R}_A \neq 0$ and $w \neq -1$. This requirement is incompatible with the simultaneous fulfillment of Hayward’s unified first law and the fundamental thermodynamic equation, as discussed in Sec.~\ref{funda_sect}. Therefore, strictly speaking, $(T_A, S)$ and $(P, V)$ are not conjugate thermodynamic variables, which also obscures their thermodynamic meanings. We expect this inconsistency to persist in other modified gravity theories as well. To overcome this obstruction, and using Euler–scaling arguments, we use an alternative phase space in which the cosmological horizon is a quasi-homogeneous system and the GUP deformation parameter is treated as a thermodynamic variable. This approach reveals a phase structure and microscopic thermodynamic behavior closely analogous to those found in AdS black holes. Moreover, our results stand in contrast to those reported in~\cite{feng2024phase}, where the phase structure is analyzed in the \( P\text{--}V \) plane and both first- and second-order phase transitions occur, depending on the value of the reduced temperature. Our analysis reveals that in the $T\text{--}U$ or $T\text{--}S$ planes the phase transition is exclusively of first-order, as demonstrated by the free–energy behavior discussed in Sec.~\ref{seccion_global}. This discrepancy stems from the fact that, in our formulation, $\beta$ is treated as an independent thermodynamic variable, whose fluctuations play a relevant thermodynamic role as shown in Sec.~\ref{GTD}. On the other hand, the analysis in \cite{feng2024phase} is based on a law of corresponding states, which suppresses any explicit thermodynamic contribution from $\beta$ and, as a result, obscures its influence on the phase structure. Still, such ambiguity is to be expected in light of the current lack of a consistent and unified framework for classifying phase transitions in gravitational systems.

Furthermore, using GTD and adopting the hypothesis that the curvature of the equilibrium space encodes the underlying thermodynamic interactions, we show that only positive values of $\beta$ can induce phase transitions in the cosmological model. Once again, and in agreement with our previous results for black holes~\cite{Ladino:2024ned, LADINO2025117031}, GTD predicts both attractive and repulsive microscopic interactions, in contrast to the purely attractive behavior inferred from the Ruppeiner scalar curvature.  Finally, we present a numerical analysis of the behavior of the GTD scalar curvature near $T_{\max}$, where we find a scaling behavior of the form $\zeta \sim 1$, independently of the dimension of the equilibrium space. This result contrasts with the exact value $\zeta = 2$ obtained in Ruppeiner geometry. Overall, these results contribute to our broader goal of understanding thermodynamic universality in gravitational systems, exploring whether the emergence of phase transitions and microstructural interactions follows a unified geometric pattern linking black holes, cosmological horizons, and other gravitational backgrounds. 

 Analyzing the fundamental equation for the GUP-FLRW universe, we find that for $\beta > 0$, there exists a minimum value for both entropy and internal energy, which may be closely related to the intrinsic uncertainty in the position and momentum operators, $\Delta x$ and $\Delta p$. In this context, the minimum entropy $S_{\min}$ can be interpreted as an effective manifestation of the minimal observable length, $\Delta x_{\text{min}} = 4\sqrt{|\beta|}$~\cite{du2022new}, which sets a fundamental limit on spatial resolution and consequently on the number of accessible microstates.   

These findings open several directions for future research. Studying the Joule–Thomson effect would allow one to analyze the cooling and heating properties of the horizon. It would be particularly interesting to extend the present analysis to FLRW cosmologies within $f(R)$ and scalar--tensor gravity theories, where the cosmological dynamics is naturally described by a non--equilibrium thermodynamics~\cite{cai2007unified}. Moreover, exploring alternative cosmological scenarios, such as models that incorporate interactions between dark energy and dark matter, may offer new perspectives on the thermodynamic description of cosmic evolution~\cite{aguilar2025interacting}.
 Finally, further exploration of higher--dimensional horizons and non--additive entropy formalisms may provide deeper understanding of the persistence of thermodynamic universality and its connection to the microscopic structure of spacetime.

\section*{Acknowledgments}
CRF acknowledges support from Conahcyt-Mexico, grant No. 4003366. The work of HQ was supported by PAPIIT-DGAPA-UNAM, grant No. 108225, and by Conahcyt,  grant No. CBF-2025-I-253.

\end{document}